\documentclass[fleqn,usenatbib]{mnras}

\usepackage{newtxtext,newtxmath}

\usepackage[T1]{fontenc}
\DeclareRobustCommand{\VAN}[3]{#2}
\let\VANthebibliography\thebibliography
\def\thebibliography{\DeclareRobustCommand{\VAN}[3]{##3}\VANthebibliography}

\usepackage{graphicx}	
\usepackage{amsmath}	
\usepackage{float}
\usepackage{nccmath}
\usepackage[subnum]{cases}
\usepackage{multirow}
\defcitealias{Murray2006}{MC2006}

\title[Neptune's Random Walk Migration]{Randomness and Retention: Using Weak Mean Motion Resonances to Constrain Neptune's Late-Stage
Migration}

\author[A. Hermosillo Ruiz et al.]{
Arcelia Hermosillo Ruiz,$^{1}$\thanks{E-mail: arcelia@ucsc.edu}
Harriet C.P.~Lau,$^{2}$
Ruth Murray-Clay$^{1}$
\\
$^{1}$Department of Astronomy and Astrophysics, University of California Santa Cruz\\
$^{2}$Department of Earth, Environmental, and Planetary Sciences, Brown University
}

\date{Accepted XXX. Received YYY; in original form ZZZ}
\pubyear{2023}
 
\begin{document} 

\label{firstpage}
\pagerange{\pageref{firstpage}--\pageref{lastpage}}
\maketitle

    \begin{abstract} 
    Planet-planetesimal interactions cause a planet to migrate, manifesting as a random walk in semi-major axis. In models for Neptune’s migration involving a gravitational upheaval, this planetesimal-driven migration is a side-eﬀect of the dynamical friction required to damp Neptune’s orbital eccentricity. This migration is noisy, potentially causing Trans Neptunian Objects (TNOs) in mean motion resonance to be lost. With Nbody simulations, we validate a previously-derived analytic model for resonance retention and determine unknown coeﬃcients. We identify the impact of random-walk (noisy) migration on resonance retention for resonances up to fourth order lying between 39 au and 75 au. 
    Using a population estimate for the weak 7:3 resonance from the well-characterized Outer Solar System Origins Survey (OSSOS), we rule out two cases: (1) a planetesimal disk  distributed between 13.3 and 39.9 au with $\gtrsim$ 30 Earth masses in today's size distribution and $T_{\rm mig} \gtrsim$ 40Myr and (2) a top-heavy size distribution with $\gtrsim$2000 Pluto-sized TNOs and $T_{\rm mig} \gtrsim$10Myr, where $T_{\rm mig}$ is Neptune's migration timescale. 
    We find that low-eccentricity TNOs in the heavily populated 5:2 resonance are easily lost due to noisy migration. Improved observations of the low-eccentricity region of the 5:2 resonance and of weak mean motion resonances with Rubin Observatory’s Legacy Survey of Space and Time (LSST) will provide better population estimates, allowing for comparison with our model’s retention fractions and providing strong evidence for or against Neptune’s random interactions with planetesimals.
    \end{abstract}  
  

\section{INTRODUCTION}  
We are approaching an era of unprecedented well-characterized data for the outer solar system with the Vera C. Rubin Observatory's Legacy Survey of Space and Time (LSST). In preparation, we revisit a previously-derived analytic model for the random walk of a planet, in semi-major axis, during planetesimal-driven migration. This model has been used to constrain the Solar System's planetesimal disk and early planetary migration using the observed population of trans-Neptunian objects (TNOs) in 3:2 first-order mean motion resonance (MMR) with Neptune \citet{Murray2006} (hereafter \citetalias{Murray2006}). We look ahead to LSST data which will observe tens of thousands of TNOs, providing enough detections and well-characterized completeness to extend this analysis across a broader range of resonances. In particular, the survey will allow us to use the fact that higher-order (weaker) resonances lose objects more easily to provide stronger constraints on the final, planetesimal-driven stage of Neptune's migration.

Over the last 20 years, many theoretical models have been developed to reproduce the orbital structure and relative populations of  TNOs  in mean motion resonances and in the classical and scattered regions of trans-Neptunian space. While none have succeeded in producing all features of this observed structure, they each have their virtues and approaches in solving the puzzle of the early Solar System's dynamical evolution (see reviews by \citealt{Nesvorny2018,Malhotra2019,Morbidelli2020,Gladman2021}). 
Early studies of TNO dynamics considered the outward migration of Neptune due to angular momentum exchange with a residual disk of planetesimals \citep{Fernandez1984}. This outward migration, later coined ``smooth migration", is efficient at capturing TNOs into mean motion resonances \citep{Malhotra1993,Malhotra1995,Hahn2005} but overpopulates these regions in semi major axis space over the classical region between the 3:2 and 2:1 mean motion resonances with Neptune. In particular, TNOs in the 3:2 resonance are observed to be $\simeq 2-4$ times less numerous than those in the hot classical belt (classical objects with high eccentricity and inclination) \citep{Gladman2012}, but \citet{Hahn2005} ``smooth migration" model produces a 3:2 MMR population $\sim$1.5 times more numerous instead. 
\cite{Nesvorny2016} finds that "graininess" introduced by placing 10-40\% of the planetesimal disk into Pluto-sized planetesimals overcomes this difficulty, producing a ratio between the hot classicals and 3:2 resonant population between 3-4 (see also \citet{Kaib2016,Lawler2019}).  This graininess produces a random walk component in semi major axis for Neptune, similar to the model discussed in this paper.  Neptune's random walk  causes resonances to lose previously captured TNOs, adding to the hot classical population. While they solve this resonance overpopulation problem, their Nbody simulation does not reproduce the orbital element distributions of the hot classical population.  In their best two migration simulations, the original disk contains 1000 planetesimals, each with either the mass of Pluto or twice the mass of Pluto. 

Alternatively, evidence of a  late heavy bombardment on Earth's Moon inspired models suggesting that the early Solar System may have had a giant planet instability early in its lifetime \citep[][and references therein]{Tsiganis2005,Pike2017,DeSousa2020}. In these instability models, Neptune is launched onto an eccentric orbit, requiring a short period of planetesimal-driven migration at the end to damp its eccentricity to its current value \citep[][]{Levison2008,Dawson2012,Wolff2012}.  
\citet{Balaji2023} found that a planetary instability that scatters planetesimals into semi major axis and eccentricity parameter space around the 3:2 MMR reproduces the semi major axis, eccentricity, and resonant angle distributions of the observed 3:2 population, but not the libration amplitude distribution. The mismatch with the libration amplitude distribution is modest enough that it may result from the addition of transient sticking \citep[][]{Lykawka2007,Yu2018} or from a brief late-stage planetesimal-driven migration, which was not rigorously modeled in their work.\footnote{The cumulative distributions of the resonant angle did not differ greatly from the libration amplitude distribution in their work; rather the statistical comparison was less constraining for the resonant angle because it accounts for its periodic nature. In addtion, the two distributions are biased by observations differently, such that the block longitudes relative to Neptune place limits on the libration amplitude magnitudes and range of phi that are observable \citep{Volk2016}. } 
The model developed in this paper applies both to scenarios in which dynamical interactions with planetesimals are the primary driver of migration and to scenarios in which a large-scale dynamical upheaval occurred among the giant planets, followed by a shorter epoch of planetesimal-driven eccentricity damping and migration.

In this work, we quantify the efficiency of retaining (and thus, losing) planetesimals in first- to fourth-order MMR with Neptune during an epoch of noisy planetesimal-driven planetary migration.
We evaluate how various planetesimal size distributions affect migration to constrain the amount of mass in large objects and typical eccentricity of planetesimals in the early disk as well as the planets' migration timescale. Simulating fully-realistic planetesimal-driven migration and resonance retention requires billions of massive planetesimals with a continuous mass distribution, since resonance retention is a function of planetesimal size distribution.
Computational limitations hinder such simulations; for example, \citet{Nesvorny2016} used $\sim 5000$ CPU days with 2000 particles and a size distribution represented by a delta function. To overcome this challenge, \citetalias{Murray2006} analytically model the planet-planetesimal encounters in analogy to Brownian motion, where the migrating planet's semi-major axis fluctuates about its mean value. 
As demonstrated in \citetalias{Murray2006}, the random component of the migration is dominated by planetesimal sizes with a maximum number density times mass squared, which is generally at the high-mass end of the planetesimal size distribution even though planetesimals with the largest sizes do not dominate the total mass. They conclude that migration was likely smooth enough to retain TNOs in first-order resonances for planetesimal disks with the bulk of the mass in bodies having sizes of order $\sim 1$ km and maximum body size of 1000 km.

In this study, we revisit and extend the work of \citetalias{Murray2006} with two aims in mind. First, we wish to assess the validity of planetary migration as a form of Brownian motion by comparing results from the analytical model of \citetalias{Murray2006} with numerical simulations.  It is important that as the planet interacts with planetesimals, it performs a random walk in semi-major axis.  We also wish to determine, by simulation, the unknown coefficients that arose during \citetalias{Murray2006}'s order-of-magnitude derivations of the change in semi-major axes of the planet and planetesimals perturbing each other during gravitational encounters.

Second, we wish to apply the theoretical expressions established in \citetalias{Murray2006} (i.e. total random walk of a planet, retention fraction in MMR) to weaker resonances and for a wider range of disk conditions and planetesimal size distributions. While the relative population of different resonances depends both on capture and retention processes, we can use retention alone as a constraint because resonances observed to host large populations of TNOs must have been able to retain these bodies.  We investigate the conditions under which weak resonances can retain TNOs. 
We find the 5:2 and 7:3 resonances most promising for placing constraints on Neptune's total migration timescale and distance and the planetesimal disk's size distribution since they are weak, contain at least some observed objects, are not located in the cold classical region, and are located near each other in semi-major axis.  The 5:2 has a large observed population, comparable to that in the 3:2 \citep{Gladman2012,Volk2016}, while the 7:3 has an observed population that is present but small \citep{Gladman2012,Crompvoets2022}. Neptune's migration is limited to an amount that would not produce a total loss to the weakest 7:3 population and allows retention of many objects in the 5:2. 
Comparing population estimates for resonances observed by the Outer Solar System Origins Survey (OSSOS; \citealt{Bannister2018}) with retention fractions produced by our model allows us to comment on the impact Neptune's noisy migration may have had on the resonances. If Neptune's noisy migration had a large impact, the mean motion resonances would preserve the retention fraction pattern found in our work. We carry out this comparison for an assumed initial surface denisty profile, assuming two commonly discussed methods for filling the mean motion resonances.

In Section \ref{secmodel} we introduce the analytical theory describing the random walk component of planetary migration. We then confirm that N-body integrations produce a change in the planet's semi-major axis that matches analytical theory. In Section \ref{applofan} we apply the analysis to all resonances (3:2, 11:7, 8:5, 5:3, 7:4, 2:1, 7:3, 5:2, 3:1, 4:1) and calculate the timescales for which objects in resonance will be lost and their resulting retention fractions. In Section \ref{sizedist} we use two end-member size distributions of the original disk during Neptune's migration to constrain the amount of largest planetsimals in the disk. In Section \ref{sec:obs}, we use the relative retention probabilities to interpret relative resonance populations constrained by OSSOS. In Section \ref{sec:discussion}, we discuss future work and additional considerations.
Finally, in Section \ref{sec:summary}, we summarize our results.

\section{analytical model and numerical verification}\label{secmodel}

Planetesimal-driven migration results from a superposition of discrete scattering events due to multiple encounters between the planet and planetesimals. This type of migration can be separated into average and random components. The average component arises from the effects of dynamical friction in a uniform sea of planetesimals, modified by asymmetries in mass, eccentricity, and number of planetesimals interior and exterior to the planet's orbit.  For a single planet, dynamical friction causes the planet to lose orbital energy and move toward the star.  In the solar system, global transfer of planetesimals between the giant planets typically leads Jupiter to migrate inward while Saturn, Uranus, and Neptune migrate outward \citep{Fernandez1984}.  In this work, we remain agnostic about the planet's average migration. The random component of migration is due to noise that arises from variations in the orbital elements of planetesimals interacting with the planet and variations in the number of encounters. Variations in the mass of planetesimals also affect migration, such that tens of thousands of low-mass planetesimals will produce less noise than thousands of high-mass planetesimals. We treat this fact by applying the most massive planetesimal mass to our analysis in this paper. This noise can cause resonances to lose previously captured TNOs, which cannot adjust to abrupt changes in the planet's orbit that occur on timescales less than the resonant libration period.

In Section \ref{sectheory} we introduce the theory of \citetalias{Murray2006}, which provides order-of-magnitude expressions describing the random component of migration, which we refer as a random walk for the remainder of the paper. We then turn to numerical simulations to verify selected fundamental equations that predict the change in the semi-major axis of a planet ($a_{\rm p}$) due to interactions with single (Section \ref{secsingleverify}) and multiple (Section \ref{secmanyverify}) planetesimals.
Throughout this section, we choose planet and planetesimal parameters to facilitate numerical simulation. These choices, which are unrealistic for the young circumsolar disk, allow us to verify our analytic expressions, which can then be applied confidently under realistic conditions, such as those considered in Section \ref{applofan}.
For all numerical simulations, we use the N-body integrator \emph{Mercury}, version 6.2 \citep{Chambers1999} with the Bulirsch Stoer integrator. 


\subsection{Theoretical Introduction}\label{sectheory}

As a planet and planetesimal approach each other, the planet will perturb the planetesimal given its larger size and mass. The impulse of the planet on the planetesimal will change its velocity, and the amount depends on the mass of the planet, $M_{\rm p}$ how close they are to each other in semi-major axis, $x$, and the encounter time, $\Delta t$.  This change in velocity alters both the angular momentum and the energy of the planetesimal. \citetalias{Murray2006} make use of the Jacobi integral to disentangle these changes and determine the planetesimal's new orbital energy. Given that energy is a function of semi-major axis, we can then solve for the change in semi-major axis, $\Delta a$, of the planetesimal. The perturbation on the planetesimal exerts a back reaction onto the planet and due to conservation of energy of the system, we can find the change in the planet's semi-major axis, $\Delta a_p$, as a result of the encounter.

Realistically, a planet doesn't experience single encounters solely, for it is embedded in a sea of planetesimals.
The planet will undergo a large number of encounters over a given encounter time, and the cumulative impact of each individual shift will produce a random step in the planet's semi-major axis. Derivations in \citetalias{Murray2006} assume that the number of planetesimals encountering the planet over a given dynamical time (approximately an encounter time) is random and follows a Poisson distribution, with the average determined by the disk's surface density.  
Planet-planetesimal encounters are random and occur on timescales shorter than the libration period of a resonant object (i.e. $\lesssim 10^2$ yrs versus $\sim 10^4-10^5$ yrs). 

Objects initially in resonance have the potential to be lost since the semimajor axis of a resonance will take a random step along with the planet, and the object might not have enough time to adjust to remain in resonance.
To avoid loss, the total cumulative change in the planet's semi-major axis due to the random walk cannot exceed half of a resonant libration width,  $\delta a_{\rm p,lib}$. We comment on the relationship between each individual random walk step, the cumulative distance of the random walk (which is superimposed on the average migration), and the total distance migrated in Sections \ref{secmanyverify} and \ref{secpkeep}.  For a resonant planetesimal with an eccentricity of $e_{\rm res}$, 

\begin{equation}\label{alibeqn}
\delta a_{\rm p,lib} \approx 2\mathcal{C}_{\rm lib}a_{\rm p} \left(\frac{M_{\rm {p}}e_{\rm res}^{J_{3}}}{M_{*}}\right)^{1/2} \,\, ,
\end{equation}

\noindent where $\mathcal{C}_{\rm lib}$ is a constant that depends on the resonance, $J_{3}$ is the order of resonance, $a_{\rm p}$ and $M_{\rm p}$ are the semimajor axis and mass of the planet, respectively and $M_*$ is the mass of the Sun \citep[for an expository derivation, see the textbook by][]{Murray1999}.\footnote{Comparing the analytical libration width for the 3:2 MMR with the numerical width of the resonance in panel 1 of Figure 2 in \citet{Balaji2023} produced comparable results.}  The libration width, $\delta a_{\rm p,lib}$,   
depends on the resonance, the eccentricity of the planetesimal, and  the order of the resonance, such that loss is not uniform across the resonances.  Rather, it is preferential: different mean motion resonances will lose more or fewer planetesimals during planetary migration, making these population relative-loss ratios a great tracer for planetesimal-driven migration.

Before proceeding to consider the distance of a planet's random walk, first we introduce a few equations and define variables.  A planet's Hill radius is 
\begin{equation}
R_{\rm H} = a_{\rm p} \left( \frac{M_{\rm p}}{3M_*} \right)^{1/3} \, \, ,
\end{equation}
its Hill eccentricity is
\begin{equation}
e_{\rm H} \equiv R_{\rm H}/a_{\rm p} \,\, ,
\end{equation}
and its Hill velocity is
\begin{equation}
v_{\rm H} \equiv R_{\rm H}\Omega_{\rm p} \,\, ,
\end{equation}
where $\Omega_{\rm p} = (GM_*/a_p^3)^{1/2}$ is the orbital angular velocity of the planet.  The random velocity of a planetesimal, $u$, with an eccentricity of $e$ is
\begin{equation}\label{eqn-u}
u \sim e\Omega a \, \, ,
\end{equation}
where $\Omega$ is the planetesimal's orbital angular velocity and $a$ is its semimajor axis. Equation (\ref{eqn-u}) approximates the difference in velocity between an eccentric and a circular orbit at the same semi-major axis, often called the epicyclic velocity. The eccentricity and random velocity are often used interchangeably in dynamics.

We introduce a dimensionless number of order unity, $\mathcal{M}$, which serves to parameterize the surface density of planetesimals in the disk, $\Sigma_m$, such that when integrated over the surface, the total mass of the disk is $\mathcal{M}M_{\rm p}$ distributed among planetesimals of mass $m$. The surface density is parameterized for one mass $m$ because a single mass (typically the largest planetesimals) generally produce the most noise \citepalias{Murray2006}.
The mass contained in the disk is uniformly distributed between $a_{\rm d}/2$ to $3a_{\rm d}/2$, where $a_{\rm d}$ is the semi major axis at the middle of the disk, and so 

\begin{equation}\label{eqn-sigmam}
\Sigma_m = \frac{\mathcal{M}M_{\rm p}}{2 \pi a_{\rm d}^2} \,\, .
\end{equation}

\noindent We keep $a_d = 26$au constant throughout this paper to keep a consistent translation between $\mathcal{M}$ and $\Sigma_m$. We note that the real planetesimal disk need not have this simplified global structure.  
The random walk of the planet depends primarily on the surface density local to the planet (i.e. within several $R_H$) since that will dictate the encounter rate.  We use Equation~(\ref{eqn-sigmam}) to intuitively relate this local $\Sigma_m$ to the amount of solid mass needed to form the planet. 
We define $x \equiv a - a_{\rm p}$ where $\mathcal{R}$ is this distance normalized by $R_{\rm H}$ such that

\begin{equation}\label{Requation}
\mathcal{R} \equiv \frac{|x|}{R_{\rm H}} \, \, .
\end{equation}

\subsection{Single encounters}\label{secsingleverify}   
We separate encounters between the planet and a planetesimal into non-crossing and crossing depending on whether the orbits of the two bodies cross one another.
In a non-crossing orbit, the planet and planetesimal are far enough from each other, such that $|x| > ae$. For these values of $x$, the relative speed of the planetesimal is dominated by the Keplerian shear velocity $\sim$$\Omega_{\rm p} x$, such that the encounter time is approximately a dynamical time, $\Omega_{\rm p}^{-1}$. 
The expected change in $a_{\rm p}$ for a non-crossing orbit is

\begin{subnumcases}{\label{NCeqn} \Delta a_{\rm p, n} \sim}
- \mathcal{C}_1 \frac{mM_{\rm p}}{M_{*}^2} \frac{a_{\rm p}^6}{x^5}, \label{NC1} \\ 
	&if $R_{\rm H} \lesssim x  \lesssim  (v_{\rm H}/u)^{1/2}R_{\rm H}$; \nonumber \\
\mp \mathcal{C}_2\frac{m}{M_{*}} \frac{a_{\rm p}^4}{x^3} e, \label{NC2} \\ 
	&if $x  \gtrsim   (v_{\rm H}/u)^{1/2}R_{\rm H}$, \nonumber
\end{subnumcases}

\noindent where $\mathcal{C}_1$ and $\mathcal{C}_2$ are order-unity numerical factors that were not determined during derivation in \citetalias{Murray2006}.

\begin{figure}
    \includegraphics[width=3.3in]{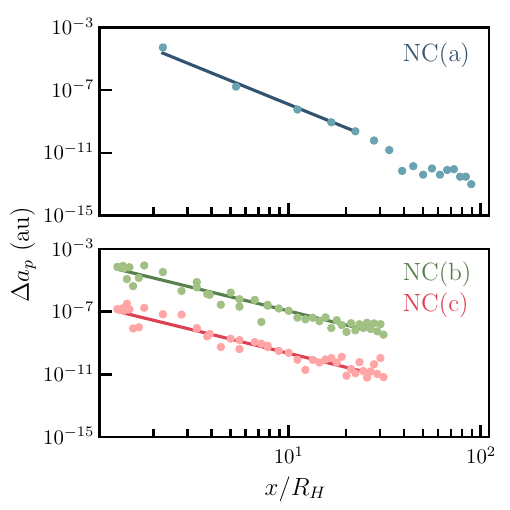}
    \caption{ Numerical simulations (circles) validate analytical Equation (\ref{NC1},\ref{NC2}) using $\mathcal{C}_1 = 6.27$ and $\mathcal{C}_2 = 2.5$ (lines) for the change in a planet's semi-major axis, $a_{\rm p}$, during a non-crossing encounter. We fit the lines to the simulations with initial $x/\mathcal{R}_H \leq 25$ since beyond that, the approximation that $x = a - a_{\rm p} \ll 1$ begins to lose validity.  We perform three sets of simulations, all with 
   planetary mass $M_{\rm p} = 10^{-5}M_{*}$ and semi-major axis $a_{\rm p} = 30$AU.  A planetesimal of mass $m$ and eccentricity $e$ starts with semi-major axis $a_{\rm p} + x$. NC(a) (top panel) verifies Equation (\ref{NC1}) with $e = 0$ and $m = 5 \times 10^{-10}M_{*}$.  NC(b) and NC(c) (bottom panel) verify Equation (\ref{NC2}) with $e = 0.01$ for masses $m = 5 \times 10^{-10}M_{*}$ (b) and $m = 10^{-12}M_{*}$ (c).} 
    \label{ncfig}
\end{figure}

\begin{figure}
\includegraphics[width=3.3in]{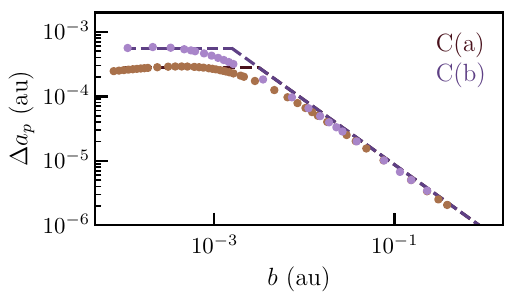}
\caption{Simulations (circles) match analytical Equations (\ref{capeqn}) and (\ref{beqn}) (lines) for the change of a planet's semi-major axis during a crossing encounter when setting $\mathcal{C}_3 = 2.5$. For both simulations, planetesimal semi-major axis, eccentricity, and mass are $a$ = 25AU, $e = 0.4$, and $m = 10^{-10}M_*$, respectively. The planet starts at semi-major axis $a_{\rm p} = 28$ au, with eccentricity $e = 0$ and masses $M_{\rm p}= 10^{-5}M_*$ for simulation set C(a) and $M_{\rm p}=5 \times 10^{-6}M_*$ for C(b). At impact parameters, $b$ less than $b_{\rm min}$ (Eq. \ref{bmineqn}), $\Delta a_{\rm p}$ is constant, as predicted by the maximum values given by Equation (\ref{capeqn}).
Beyond that, $\Delta a_{\rm p}$ falls off inversely proportionally to $b$, as in Equation (\ref{beqn}).}
\label{Cfig} 
\end{figure}

To find $\Delta a_{\rm p, n}$ numerically, we run simulations with the Sun, a planet, and a planetesimal. For the simulation to begin with the minimum amount of influence experienced by each body on the other, the planet and planetesimal begin at opposite sides of the Sun. The simulation runs for a synodic period so that a close encounter will occur and enough time after the encounter is left for the change in the semi-major axis to settle to its final value. We test three non-crossing orbits, verifying the two regimes in Equation (\ref{NCeqn}), as seen in Figure \ref{ncfig}. All simulations have $M_{\rm p} = 10^{-5}M_{*}$, $a_{\rm p} = 30$AU, and planetary eccentricity $e_p = 0$.
Simulation set NC(a) has $e = 0$, appropriate for Equation (\ref{NC1}), while NC(b) and (c) have $e = 0.01$, appropriate for Equation (\ref{NC2}).  Planetesimals are given mass $m = 5 \times 10^{-10}M_{*}$ for NC(a) and (b) or $m = 10^{-12}M_{*}$ for (c) and have initial semi-major axes $a_{\rm p} + x$ for a range of values of $x$ shown in Figure \ref{ncfig}. 
We determine the numerical factors $\mathcal{C}_1$ and $\mathcal{C}_2$ by fitting the simulation output to Equations (\ref{NC1}) and (\ref{NC2}) with the linear least-squares method in log space. We perform the fit to simulation outputs with planetesimals initialized at $x/\mathcal{R}_H \le 25$ since beyond that the approximation that $x = a - a_{\rm p} \ll 1$ begins to break and the slope for $\Delta a_{\rm p}$ begins to flatten out. The three simulations had different best-fit parameters of $C_1 = 6.27$, $C_2 = 2.17$, and $C_2 = 2.78$, respectively. For simulation sets (b) and (c), we expect the same coefficient value, so we adopt the average of the latter and $\mathcal{C}_2 = 2.5$.

In a crossing orbit, $|x| < ae$, such that $u > \Omega x $
so the relative velocity of the planetesimal is dominated by its epicylic velocity and its encounter time is less than $\Omega^{-1}$.
The impact parameter $b$ can be rather different from $|x|$, and $b$ lies in the range
\begin{equation}
b_{c} < b \lesssim ae \,\, ,
\end{equation}
where $b_{c}$ is the impact parameter below which a collision occurs.  This wide range can cause a wide spectrum of results for $\Delta a_{\rm p,c}$ (where $\Delta a_{\rm p,c}$ is the change of $a_{\rm p}$ due to a crossing encounter) 
so \citetalias{Murray2006} limit themselves to predict the maximum value of $|\Delta a_{\rm p,c}|$ possible.  Weaker encounters contribute to the random walk of the planet at a comparable rate, generating a Coulomb logarithm, which we fold into the coefficient calculated in Section \ref{secmanyverify}.  The maximum $|\Delta a_{\rm p,c}|$ over an encounter is
\begin{equation}\label{capeqn}
\max\left| \Delta a_{\rm p,c}\right| \sim \mathcal{C}_3 \frac{m}{M_{\rm p}}a_{\rm p}e \,\, ,
\end{equation}
generated when the impact parameter satisfies
\begin{equation}
b \lesssim GM_{\rm p}/u^2 \,\, ,
\label{bmineqn}
\end{equation}

\noindent Once $b$ does not satisfy Equation (\ref{bmineqn}) we expect $|\Delta a_{\rm p,c}|$ to decrease from the value in Equation (\ref{capeqn}), following the relationship

\begin{equation}
\left| \Delta a_{\rm p,c}\right| \sim \mathcal{C}_3 \frac{Gm}{bae \Omega^2} \propto \frac{1}{b} \,\, .
\label{beqn}
\end{equation}

As before, we find $\Delta a_{\rm p,c}$ numerically with simulations containing the Sun, a planet, and a planetesimal. 
To ensure the planet and planetesimal are in a crossing orbit, we begin with the planet on a circular orbit with $a_p = 28$au and place an eccentric planetesimal in the same plane.  We set the planet's and the planetesimal's longitude of ascending node and argument of pericenter to be zero.  Under these conditions, a collision occurs if the true anomalies of both objects coincide simultaneously at a place where the two orbits cross.  Since the planet's orbit is circular, this crossing must occur at a distance from the sun $r = a_{\rm p}$, while the planetesimal's orbit satisfies $r = a(1-e^2)/(1+e\cos{f})$, so that an encounter occurs at angle $f_{\rm enc} = \cos^{-1}\left[(a(1-e^2)-a_p)/(a_p e)\right]$.  We begin the simulation at time $\delta t$ = 2.5 years before the collision, when the mean anomalies of the objects are smaller by $\Delta \mathbb{M}_{\rm p} = \Omega_{\rm p} \delta t$ for the planet and $\Delta \mathbb{M} = \Omega \delta t$ for the planetesimal. Since it is on a circular orbit, the planet's initial mean anomaly is $\mathbb{M}_{\rm p} = f_{\rm enc} - \Delta \mathbb{M}_{\rm p}$.  The planetesimal's initial mean anomaly is $\mathbb{M} = \mathbb{M}_{\rm enc} - \Delta \mathbb{M}$, where we calculate the mean anomaly at encounter, $\mathbb{M}_{\rm enc}$, from $f_{\rm enc}$ using Kepler's equation $\mathbb{M}_{\rm enc} = E - e\sin{E}$ for the eccentric anomaly $E = \tan(f_{\rm enc}/2)$. Because we want close encounters rather than perfect collisions, we add a small random value of order $10^{-4}$ radians to the planetesimal's initial mean anomaly  and, for each simulation, we record the distance of closest approach, $b$.       

We run two sets of simulations with various impact parameters, $b$, to verify Equations (\ref{capeqn}) and (\ref{beqn}) (Figure \ref{Cfig}).
The planetesimals in both simulations have $a= 25$ au, $e= 0.4$, and $m=10^{-10}M_{*}$. 
Simulation set C(a) has $M_p=10^{-5} M_*$ whereas C(b) has $M_{\rm p}=5 \times 10^{-6} M_{*}$, thus highlighting the fact that the planet's mass determines the maximum value of $|\Delta a_{\rm p,c}|$ (Equation \ref{capeqn}) but not its value at large impact parameters (Equation \ref{beqn}). Our simulations validate the equations for crossing orbits with $\mathcal{C}_3 = 2.5$  (Figure \ref{Cfig} dashed lines).

\subsection{Multiple encounters}\label{secmanyverify}  
The broad-scale level of noise for a planet in a sea of planetesimals relies on the single encounters discussed above, as the cumulative effects are essentially a randomized distribution of those events. When multiple planetesimals encounter the planet simultaneously, the planet and planetesimals exchange angular momentum, and the planet migrates due to the back reaction it experiences. We now consider this scenario, where the change of $a_{\rm p}$ (i.e. its random walk in semi-major axis) is well modeled by Brownian motion. 

\citetalias{Murray2006} derive the expected random walk, $\Delta a_{\rm p,T}$, of a planet in a disk of planetesimals given by

\begin{equation}
    \Delta a_{\rm p,T} = (\dot{\overline{N}}T)^{1/2} \Delta a_{\rm p}
    \label{randomwalkplanet}
\end{equation}

\noindent where $\dot{\overline{N}}$ is the mean encounter rate, $T$ is the migration duration, and $\Delta a_{\rm p}$ is the change in semi-major axis of the planet due to a single encounter as shown in the previous two subsections. For a non-crossing orbit, $u<v_H$, the speed of the planetesimals relative to the planet is determined by Keplerian shear, and impact parameter $b\sim x$ resulting in a mean encounter rate, 

\begin{equation}
    \dot{\overline{N}} \sim \frac{\Sigma_m}{m}\Omega_p x^2
    \label{meanNdotnoncross}
\end{equation}

\noindent Planetesimals in crossing orbits encounter the planet with super-Hill velocities ($u>v_H$) at impact parameters $b\lesssim GM_p/u^2$ resulting in a mean encounter rate, 

\begin{equation}
    \dot{\overline{N}} \sim \frac{\Sigma_m \Omega_p}{m u} \Big(\frac{GM_p}{u^2}\Big)^2 u
    \label{meanNdotcross}
\end{equation}

\noindent where $u \sim a_p e \Omega_p$.

We summarize Equation \ref{randomwalkplanet} for the non-crossing and crossing cases as
\begin{subnumcases}{\label{RMSeqn}  \Delta a_{\rm p,T} \sim}
\mathcal{CR}^{-4} \left( \frac{\mathcal{M}m}{M_{\rm p}} \right)^{1/2} \frac{a_{\rm p}}{a_{\rm d}}e_{\rm H}v_{\rm H} \left( \frac{T}{\Omega_{\rm p}} \right)^{1/2}, \label{RMSeqn1}\\ \nonumber \\ \nonumber
&\hspace{-28mm}if $e<e_{\rm H}/\mathcal{R}^2$; \\
\mathcal{CR}^{-2} \left( \frac{\mathcal{M}m}{M_{\rm p}} \right)^{1/2} \frac{a_{\rm p}}{a_{\rm d}}ev_{\rm H} \left( \frac{T}{\Omega_{\rm p}} \right)^{1/2}, \label{RMSeqn2}\\ \nonumber \\ \nonumber
&\hspace{-28mm}if $e_{\rm H}/\mathcal{R}^2<e<\mathcal{R}e_{\rm H}$; \\
\mathcal{C} \left( \frac{\mathcal{M}m}{M_{\rm p}} \right)^{1/2} \frac{a_{\rm p}}{a_{\rm d}}
\frac{e_{\rm H}^2}{e}v_{\rm H} \left( \frac{T}{\Omega_{\rm p}} \right)^{1/2}, \label{RMSeqn3} \\ \nonumber \\ \nonumber
&\hspace{-28mm}if $e>\mathcal{R}e_{\rm H}$,
\end{subnumcases}
where $\mathcal{C}$ is a constant that encapsulates all factors of order unity that arose during derivations, and $e$ is the eccentricity of the planetesimals. The non-crossing orbit scenario takes two forms (Equations \ref{RMSeqn1} and 
\ref{RMSeqn2}) since $\Delta a_{\rm p}$ will vary depending on how $|\Delta e|$ compares to the pre-encounter $e$. Equation (\ref{RMSeqn3}) corresponds to crossing orbits. Note that Equation (\ref{RMSeqn3}) does not depend on $\mathcal{R}$, so all planetesimals with crossing orbits and the same eccentricities will affect the random walk of the planet equally.

With numerical simulations, we determine the unknown coefficient $\mathcal{C}$ (where $\mathcal{C} > 1$ but not $\gg 1$) and verify the total random walk of the planet, $\Delta a_{\rm p,T}$, due to continuous interactions with planetesimals.We run two sets of 150 simulations, each lasting one million years with the Sun, a planet, and 9000 planetesimals.
We do not know the typical eccentricity of the disk of planetesimals at the time of Neptune's migration, so we initialize one set of simulations with low eccentricity planetesimals and the other with high eccentricity planetesimals, which we call D(a) and D(b), respectively. Still, these properties are set by the disk's global dynamics, and both regimes likely occur. Here, the characterization of "low" and "high" is relative to the Hill eccentricity, $e_{\rm H}$, since disks with $e > e_{\rm H}$ host planetesimals on crossing orbits even if the region within a Hill radius of the planet is dynamically cleared.

For all simulations the planet begins at 25 au with $M_{\rm p}=10^{-6}M_{*}$, so that $e_{\rm H}=0.007$. We explore the disk's low and high eccentricity regimes with simulation sets D(a) and D(b) where the initial $e = 0.001$ and $e = 0.05$, respectively.
The planetesimals are all of equal mass $m = 1.8 \times 10^{-10} M_{*}$.  Their semi-major axes $a$ are uniformly distributed between $\sim$20 (21.5) au and $\sim$30 (28.5) au for D(a) (D(b)), inclinations between 0 and the value of $e$ in radians, and remaining orbital angles are randomly generated between 0, 2$\pi$.    
Gravitational interactions between the planet and the planetesimals are simulated, but interactions between planetesimals themselves are not.
The planet and planetesimal disk have a comparable mass, so one might wonder why neglecting planetesimal-planetesimal interactions here is appropriate. We discuss the reason for this assumption in Section \ref{sec:discussion}.

\begin{figure}
	\centering
    \includegraphics[width=3.3in]{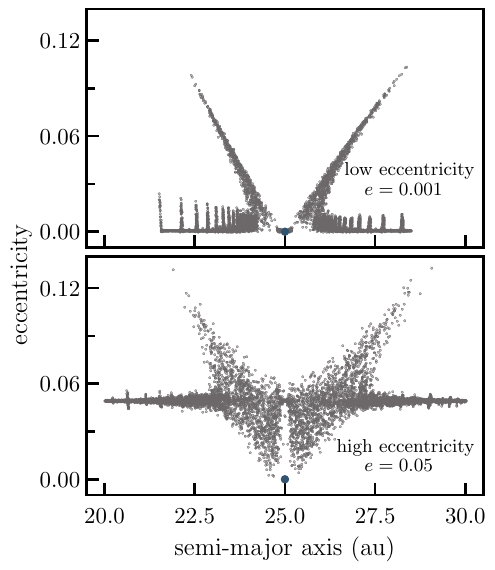}
    \caption{Typical $a-e$ snapshot after $10^6$ years for one of the 150 low-eccentricity (top) and high-eccentricty (bottom) simulations discussed in Section \ref{secmanyverify}. For these simulations, we place a planet within a disk of 9000 planetesimals with low eccentricities of 0.001 (top) and high eccentricities of 0.05 (bottom). The planetesimals (grey dots) have equal mass ($1.8 \times 10^{-10} M_{*}$) and are uniformly distributed in semi-major axis space between 20 au and 30 au. Their inclinations are between 0 and the value of their respective eccentricity in radians. The planet (blue dot) begins the simulation at 25 au with $M_{\rm p}=10^{-6}M_{*}$ and will migrate through the disk due to multiple encounters with the planetesimals. We confirm that the disk width is sufficiently wide to test for migration and that the systems have reached pseudo steady-state. Excitation at some resonances occurs (vertical lines).}
    \label{aeplots}
\end{figure}

\begin{figure}
\centering
\includegraphics[width=3.3in]{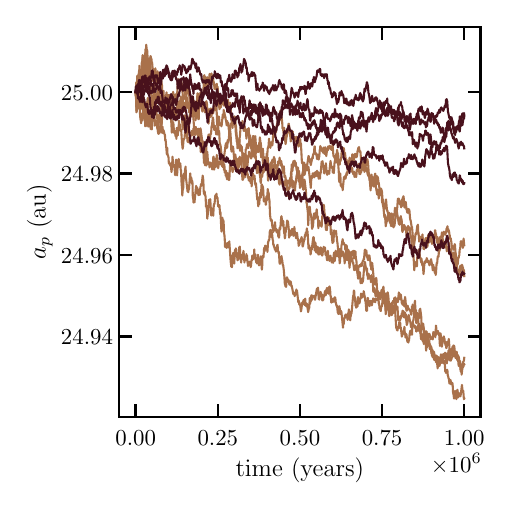}
\caption{Typical random walks in semi-major axis of the planet from 5 simualtions from the low eccentricity (light brown) and high eccentricity (dark brown) ensembles. The planet starts off at 25 au and random walks for the duration of the simulation. The trajectories resulting from migration within a disk of planetesimals follow a random walk.}
\label{randomwalkfig}
\end{figure}

For Equation (\ref{RMSeqn}) to be valid, the system must reach a pseudo-steady state by the end of the simulation, meaning over the timescale we are simulating, the distribution of particles and surface density of the disk is not changing substantially. We confirm that the disk is wide enough such that the planetesimals towards the edge of the disk are not heavily excited by the planet and the planet does not migrate outside the disk. After several thousands of years, the system converges to a stable configuration. The snapshot at $10^5$ years looks very similar to the snapshot at $10^6$ years. The one-million-year snapshot in Figure \ref{aeplots} shows the extent of influence of the planet along with the excitation of planetesimals at positions of mean motion resonances.   
In addition, we illustrate in Figure \ref{randomwalkfig} that the typical paths taken by the planet across several simulations are indeed random.   

\begin{figure}
	\centering
    \includegraphics[width=3.3in]{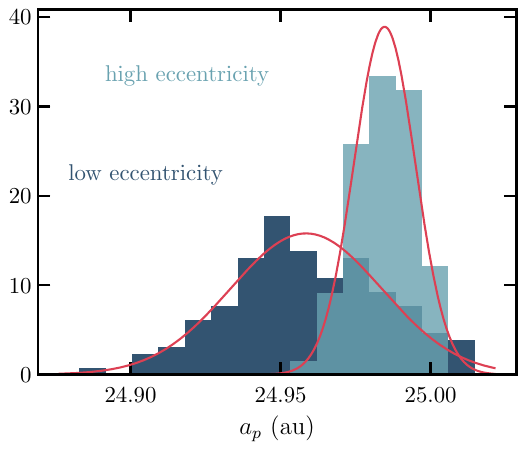}
    \caption{Histograms showing the final semi-major axis of the planet, $a_{\rm p}$, after $10^6$ years for the 150 runs performed for the low eccentricity simulation D(a) (dark blue) and high eccentricity simulation D(b) (light blue) described in Figure \ref{aeplots}. The planet's initial $a_{\rm p} = 25$ au for all runs.  The low eccentricity regime produces both a larger shift in the mean and a larger spread of final positions. The mean moves inward due to dynamical friction because our disk contains only one planet, eliminating global asymmetries which can result in outward migration. 
    The distribution of final positions is well-modeled by Gaussians (dark pink lines), which for the low eccentricity (high eccentricity) case have mean $\mu$ = 24.9585 au (24.9847 au) and standard deviation $\sigma$ = 0.0253 au (0.0103 au). This demonstrates that the planet's migration is well modeled as a diffusion process.}
    \label{histfig}
\end{figure}

We average over 150 simulations for each regime to determine the randomness of this migration process. 
At $10^6$ years, the planet's final semi-major axis for the low eccentricity (high eccentricity) case is dispersed among an ensemble of simulations with mean value $\sim 24.9585$ au (24.9847 au) and standard deviation 0.0253 au (0.0103 au). The mean semi-major axis moves towards the Sun after $10^6$ years (Figure \ref{histfig}), as expected for a simulation containing a single planet. 
In the Solar System, planetesimals scattered by the outermost giant planets are preferentially removed from Neptune's dynamical environment by interactions with Jupiter and Saturn.  As a result, Neptune migrates outward \citep{Fernandez1984}. Because the random walk arises due to Poisson noise on top of the mean number of planetesimal encounters on one or both sides of the planet \citepalias{Murray2006}, the magnitude of the random walk is not expected to depend on the direction of motion. 
In Appendix \ref{sec:appendix}, we illustrate that the average migration seen in our simulations matches that expected for a single planet in a planetesimal disk.

For the remainder of this work, we consider only the random walk component. 
We interpret the typical order-of-magnitude random walk distance $\Delta a_{\rm p,T} $ given by Equation (\ref{RMSeqn}) as the standard deviation of the planet's semi-major axis across the 150 simulations, $\sigma_{a_{\rm p, \rm T}}$.

\begin{figure}
	\centering
    \includegraphics[width=3.3in]{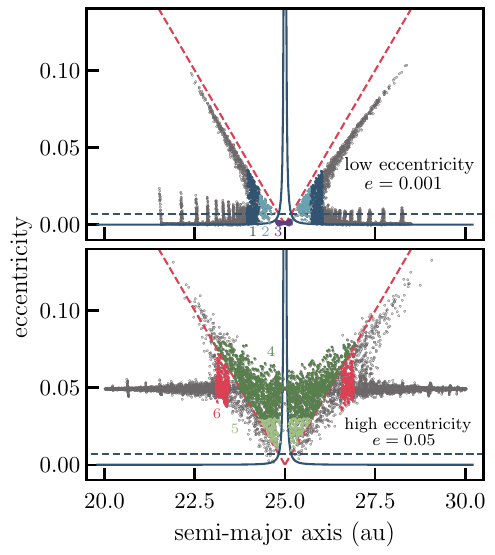}
    \caption{
    The planetesimals' relative distances to the planet, eccentricities, local disk surface density, and more, greatly affect the magnitude of the planet's random walk. We show the same simulation snapshots as Figure \ref{aeplots} to visualize the orbital differences between six regions in the disk, to probe which region produces the higher $\Delta a_{\rm p,T}$. 
    The six groups--denoted by the colors (numbers) dark blue (1), light blue (2), dark purple (3), dark green (4), light green (5), and dark pink (6)--are shown in Table \ref{groupstable} and Figure \ref{stdevfig}.  The planetesimals in group 3 have eccentricities $\lesssim e_H/\mathcal{R}^2$ (blue solid line) so Equation (\ref{RMSeqn1}) determines their impact on the planet's noisiness.  Planetesimals in groups 1, 2, and 6 have eccentricities $\gtrsim e_H/R^2$ but $\lesssim\mathcal{R}e_H$ (dark pink dashed line) so Equation (\ref{RMSeqn2}) determines their impact. Planetesimals in groups 4 and 5 occupy crossing orbits with eccentricities $\gtrsim\mathcal{R}e_H$, and their impact on the planet's random walk is determined by Equation (\ref{RMSeqn3}). The Hill eccentricity of the planet is shown for reference (blue dashed line). 
    }
    \label{singleaefig}
\end{figure}

Because Equation (\ref{RMSeqn}) depends on several orbital parameters of the disk and properties of the planet, we must disentangle which sub-population of simulated planetesimals causes the most randomness. To do that, we choose six regions of $a-e$ space, as shown in Figure \ref{singleaefig}, for the low eccentricity (D(a), top) and high eccentricity (D(b), bottom) simulation sets. 
All parameters that are the same across the six groups are described in Figure \ref{aeplots}, whereas $\mathcal{R}$, $e$, and $\mathcal{M}$ are shown in Table \ref{groupstable}. 

We calculate the surface density of both disks for each group centered at $\pm x$ au away from the planet, where $x = \mathcal{R}R_H$. 
For groups $1-3$ and 6, the groups have radial width $2R_H$ and eccentricities within the limits expressed in Equations (\ref{RMSeqn1}) and (\ref{RMSeqn2}). For groups 4 and 5, we include all planetesimals in semi-major axis space appropriate for the crossing orbits, (denoted by the dark pink dashed line in Figure \ref{singleaefig}), since Equation (\ref{RMSeqn3}) does not depend on $\mathcal{R}$. The two groups are cut off at eccentricity 0.03, and the width used to calculate the surface densities for the groups is $(25 + |x|)e$ au.
The non-dimensional surface density parameter, $\mathcal{M}$, is then determined by Equation (\ref{eqn-sigmam}) summing $\mathcal{M}$ pertaining to the inner and outer disk for each group.
With this exercise, we wish to determine which factor plays the larger role in producing a higher $\Delta a_{\rm p,T} $: eccentricity of the planetesimals or proximity to the planet. 

\begin{table}
\centering
\begin{tabular}{c c c c c c c}
\hline
 & group no. & $e$ & $\mathcal{R}$ & $\mathcal{M}$ &equation \\ \hline
\multirow{3}{*}{D(a)} & 1 & 0.001 ($e_i$) & 5 & 14.72 & \ref{RMSeqn2}  \\
 & 2 & 0.0069 ($e_{\rm H}$) & 3 & 2.56& \ref{RMSeqn2}  \\
 & 3 & 0.001 ($e_i$) & 1 & 6.75 & \ref{RMSeqn1} \\ \hline
& 4 & 0.05 ($e_i$) & 1 & 6.38& \ref{RMSeqn3}  \\
\multirow{1}{*}{D(b)} & 5 & 0.01 ($e_i$) & 1 & 4.16&  \ref{RMSeqn3} \\
 & 6 & 0.05 ($e_i$) & 10 & 8.56 &\ref{RMSeqn2}  \\
\hline
\end{tabular}
\caption{Summary of the parameters that differ across groups 1-6 (illustrated by the colored groups in Figure \ref{singleaefig}). Groups 1-3 probe the low eccentricity simulation ensemble, whereas 4-6 probe the high eccentricity simulation ensemble. The planetesimals in each group occupy non-crossing and crossing orbits, and will therefore interact with the planet differently and produce varying magnitudes of random walk distance, determined by the equation denoted in the last column.}
\label{groupstable}
\end{table}

\begin{figure}
	\centering
   
        \includegraphics[width=3.3in]{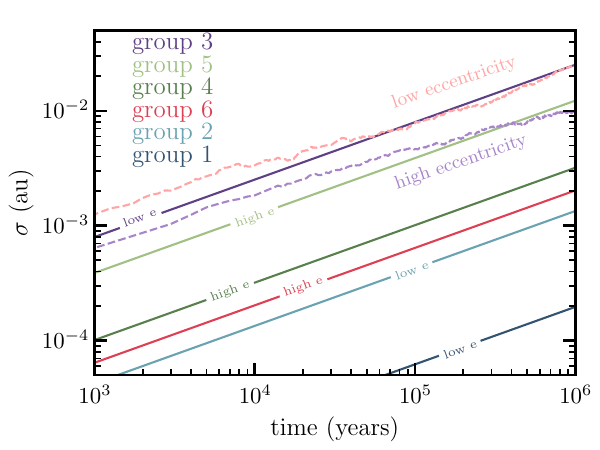}

    \caption{The numerical, expected planet's random walk, determined by the standard deviation, $\sigma$, of the distribution of semi major axes over time for simulation sets D(a) (light pink dashed line) and D(b) (light purple dashed line)  
    compared to the analytical random walk for each group shown in Figure \ref{singleaefig} and Table \ref{groupstable} (solid lines), determined by Equation (\ref{RMSeqn}). The orbital parameters and local surface density from Groups 3 (dark purple) and 5 (light green) produce the largest $\Delta a_{\rm p,T}$ for the low and high eccentricity case, respectively, highlighting the role of proximity of the planetesimals to the planet and their eccentricity. The group labels are shown in order from largest $\Delta a_{\rm p,T}$ to smallest and inline labels to clearly compare between low and high eccentricity groups.}
    \label{stdevfig}
\end{figure}

Inputting the values in Table \ref{groupstable} and Figure \ref{aeplots}, we calculate Equation (\ref{RMSeqn}) for each group as a function of time and compare to the amount of migration from simulations, as shown in Figure \ref{stdevfig}. The simulation-derived migration for low eccentricity (dashed pink line) and high eccentricity (dashed purple line) ensembles are best matched by the analytical curves with parameters from D(a) group 3 and D(b) group 5 and $\mathcal{C} = 3.5$. These groups have the largest impact on the planet's random walk (i.e. have largest values of $\Delta a_{\rm p,T}$). 
In low eccentricity case (a) groups 1 and 2 underestimate the migration by factors greater than 10, compared to group 3, so the planetesimals in group 3 clearly dominate the migration. This is likely because they are closely packed at one Hill radius from the planet at low eccentricities. For the high eccentricity case (b), groups 4 and 6 are lower by factors less than 10, compared to group 5, so we can say that planetesimals across the disk still affect migration but not as greatly as those closer to the planet with lower eccentricities.

Setting $\mathcal C$ to 3.5 for Equation (\ref{RMSeqn}) provides a good fit to $\sigma_{a_{\rm P,T}}$ for cases (a) and (b). 
Note that, particularly for the low eccentricity case, the slope of the numerical curve matches our theoretical expectation of $\Delta a_{\rm p} \propto T^{1/2}$, both near the beginning of the simulation and near the end, but the normalization changes at intermediate times.  At early times, the disk has not yet reached a pseudo steady-state, in which some regions near the disk have been cleared by scattering (see Figure \ref{singleaefig}), leading to enhanced but transient dispersion. 

For the high eccentricity case, the late time slope is less convincingly consistent with our theoretical expectation, but again, the early time slope matches.  The transition to a true pseudo steady state may simply be taking longer in this case---longer integrations would clarify this picture. Our results provide good verification of Equation (\ref{RMSeqn}) and convince us that planetary migration within a disk of planetesimals can be modelled by Brownian motion and that no cumulative effects build up, i.e each encounter is truly random.

\section{analysis applied to all resonances}\label{applofan}

With a confirmed random walk analytical model and normalized coefficients, we apply the analysis from Section \ref{secmodel} to the Solar Systems' early planetesimal disk present during Neptune's migration. The random walk of a planet is dominated by the largest planetesimals driving migration (\citetalias{Murray2006}), so we proceed to consider an average size for the largest planetesimals observed today in the Kuiper Belt \citep{Brown2008}. We use the analytics from Section \ref{secmodel} to find the characteristic times for loss from various resonances with Neptune (Section \ref{seclosstime}) and then calculate the retention fractions for these resonances (Section \ref{secpkeep}). We analyze the following characteristic loss times and retention fractions as a function of $\mathcal{M} = \mathcal{M}_{\rm LG}$, which represents the surface density of the disk in planetesimals with radius equal to 780 km and $\rho = 2 \rm g/cm^2$, resulting in a mass that is roughly one-third of Pluto's mass.  We choose this radius because it represents the average of the largest 8 observed TNOs in our solar system from \citet{Brown2008}.
The characteristic loss times and retention fractions can give us an upper limit migration timescale constraint before objects are lost from resonance. 

For our analysis, we assume an early planetesimal disk with total mass given by $\mathcal{M} = 2$, which corresponds to a total of $\gtrsim 30$ M$_\oplus$. Planetesimal disk masses between 20 to 35 $M_{\oplus}$ are typical for models of Neptune's migration \citep{Luu2002,Nesvorny2018}. Within this framework, $\mathcal{M}_{\rm{LG}} = 2$, where $\mathcal{M}_{\rm{LG}}$ is the mass in large planetesimals, would imply that 100\% of the mass in the disk is contained in large objects, while smaller values of $\mathcal{M}_{\rm LG}$ imply that only a fraction $\mathcal{M}_{\rm LG}/\mathcal{M}$ of the total mass is in these large bodies. Thus, the value of $\mathcal{M}_{\rm LG}$ for a given time is directly related to the size distribution of the planetesimal disk. In addition to $\mathcal{M}_{LG}$, we do this analysis for $\mathcal{M}_{PL}$ and $\mathcal{M}_{2PL}$ corresponding to Pluto size and twice the size of Pluto objects. 
We discuss constraints for the size distribution in relation to retention fractions of MMR during the epoch of Neptune's migration for values of $\mathcal{M}_{LG}$, $\mathcal{M}_{PL}$, and $\mathcal{M}_{2PL}$ in Section \ref{sizedist}.

\subsection{Ranking of loss times}\label{seclosstime}
\begin{table}
\centering
\begin{tabular}{ c c  c  c  c}
\hline
$p:q$ & $a_{\rm res}$ (au) & $\mathcal{C}_{\rm lib}$  & $e_{\rm res}$ & $e_{\rm res}$ reference\\ 
\hline
3:2 & 39.4 & 3.64 & 0.175 & \citet{Volk2016}\\
11:7 & 40.7 & 23.66 & 0.18 & $q$-fit\\
8:5 & 41.2 & 11.16 & 0.19 & $q$-fit\\
5:3 & 42.3 & 5.51 & 0.16 & \citet{Gladman2012}\\
7:4 & 43.7 & 9.03 & 0.12 & \citet{Gladman2012}\\
9:5 & 44.5 & 15.39 & 0.22 & $q$-fit\\
2:1 & 47.8 & 3.00 & 0.275 & \citet{Chen2019}\\
7:3 & 53.0 & 8.87 & 0.30 & \citet{Gladman2012}\\
5:2 & 54.4 & 5.26 &0.40 & \citet{Volk2016}\\
3:1 & 62.5 & 3.25 & 0.48 & \citet{Alexandersen2016}\\
4:1 & 75.7 & 3.64 & 0.57 & \citet{Alexandersen2016} \\
\hline
\end{tabular}
\caption{Mean motion resonances up to forth order lying between the 3:2 and 4:1 resonances (denoted by $p:q$), listed in order of ascending semi major axis, $a_{\rm res}$. Values of $\mathcal{C}_{\rm lib}$ are provided for reference (see \citet{Murray1999}). For resonances with a citation listed in column five, we quote the average value of eccentricities of observed resonant TNOs, $e_{\rm res}$, which may be affected by selection effects.  The remaining resonances' eccentricities were determined by fitting the data to $e_{\rm res} = 1 - q_{\rm free}/a_{\rm res}$, $q_{\rm free}$ is a free parameter we fit. The best-fit pericenter distance, $q_{\rm free}$, value was then used to set a typical $e_{\rm res}$.}
\label{resonancetable}
\end{table}

We calculate the maximum migration timescale Neptune can migrate before losing objects in resonance given that a planet
must random walk a distance less than half the libration width of the resonance in order to maintain planetesimals in a mean motion resonance ($\Delta a_{\rm p,T} < \delta a_{\rm p,lib}/2$).   
We call this timescale, the resonance loss timescale, $T_{\rm loss}$, which can also be interpreted as the migration duration for which a planet will random walk a total of more than $\delta a_{\rm p,lib}/2$, thus causing loss from resonances for given parameters of the disk.
Using Equation (\ref{alibeqn}) and Equation (\ref{RMSeqn1}), appropriate for $e<e_{\rm H}/\mathcal{R}^2$ (other regimes produce analogous results), the resonance loss timescale is
\begin{equation}\label{eqnlosstime}
T_{\rm loss} \sim \mathcal{R}^8 \left( \frac{\mathcal{C}_{\rm lib}}{\mathcal{C}} \right)^2 \frac{M_{\rm p}^2}{\mathcal{M}M_{*}m} \frac{e_{\rm res}^{J_3}}{e_H^2} \left( \frac{a_{\rm d}}{v_{\rm H}} \right)^2  \Omega_{\rm p} \, \, .
\end{equation}

\begin{figure}
\centering
\includegraphics[width=3.3in]{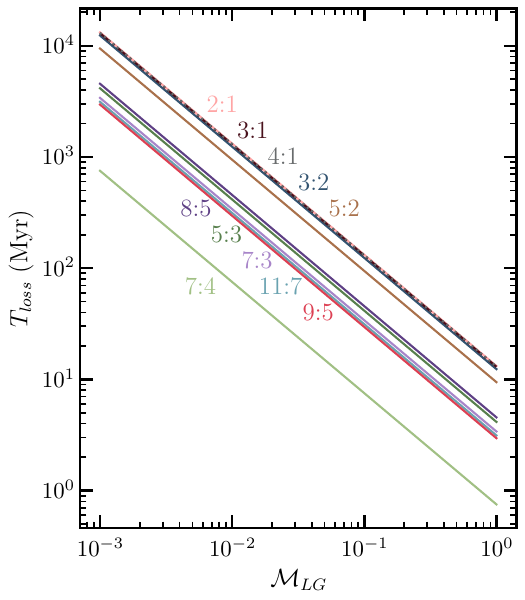}
\caption{Characteristic resonance loss time, $T_{\rm loss}$, as defined by Equation (\ref{eqnlosstime}) for the exterior mean motion resonances of Neptune that we study (see Table \ref{resonancetable}), as a function of $\mathcal{M} = \mathcal{M}_{\rm LG}$ (see Equation \ref{eqn-sigmam}).  For this calculation, $\mathcal{R} = 1$,  $e = 0.01$, $a_p = a_d = 26.6$ au, $m = 0.3M_{\rm PL}$, $M_{\rm p} = M_{\rm N}$, and $M_* = M_\odot$, where $M_{\rm PL}$, $M_{\rm N}$ and $M_\odot$ are the masses of Pluto, Neptune and the Sun, respectively, so that $R_{\rm H} \sim 0.69$ au and $e_{\rm H} \sim 0.026$. 
The eccentricity of particles in resonance, $e_{\rm res}$, is set to the typical value observed today for each resonance or a value determined by fitting the data to $e_{\rm res} = 1 - q_{\rm res}/a_{\rm res}$ for those without measurements (see Table \ref{resonancetable} and text).
}
\label{losstimefig1}
\end{figure}

Figure \ref{losstimefig1} presents $T_{\rm loss}$ for our collection of resonances, evaluated using Equation (\ref{eqnlosstime}) with $\mathcal{R} = 1$,  $e = 0.01$, $a_p = a_d = 26.6$ au, $m = 0.3M_{\rm PL}$, $M_{\rm p} = M_{\rm N}$, and $M_* = M_\odot$, where $M_{\rm PL}$, $M_{\rm N}$ and $M_\odot$ are the masses of Pluto, Neptune and the Sun, respectively, so that $R_{\rm H} \sim 0.69$ au and $e_{\rm H} \sim 0.026$.  We set $\mathcal{C} = 3.5$, the value determined in Section \ref{secmanyverify}, and plug in typical eccentricities of TNOs in MMR for $e_{\rm res}$ (see Table \ref{resonancetable}). We do this study for resonances up to 4th order that lie between the 3:2 and 4:1. For the resonances with enough observations to produce a typical eccentricity range for the TNOs, we quote their Gaussian centered or average value in Table \ref{resonancetable} based on surveys done by \citet[]{Gladman2012,Volk2016,Crompvoets2022,Bannister2018,Alexandersen2016,Chen2019,Volk2016}. We determine $e_{\rm res}$ for the resonances missing observed eccentricities by assuming the resonances possibly have a similar pericenter distribution, $q_{\rm free}$, motivated by the observed eccentricities increasing as a function of semi-major axis. Similarly stable libration zone studies in $(a,e)$ space for Neptune's exterior MMRs show that the widest part of the libration width (i.e. where more TNOs are expected to reside) increases along a line of constant pericenter \citep{Lan2019}. Fitting the data in Table \ref{resonancetable} to $e_{\rm res}=1 - q_{\rm free}/a_{\rm res}$ produced a best-fit value $q_{\rm free}=\sim 34.5$ au, resulting in appropriate $e_{\rm res}$ values equal to 0.18, 0.19, and 0.22 for the 11:7, 8:5, and 9:5 MMRs, respectively.

Because the noise due to the planet's random walk is dominated by the largest planetesimals driving migration, we show $T_{\rm loss}$ as a function of $\mathcal{M} = \mathcal{M}_{\rm LG}$. 
The MMRs requiring the longest time to lose objects (2:1, 3:1, 4:1, 3:2; Figure \ref{losstimefig1}) can be interpreted as being the strongest resonances, whereas the 7:3, 11:7, 9:5, and 7:4 are the weakest. The order in Figure \ref{losstimefig1} matches the order of $\delta a_{\rm p,lib}$ for the values in Table \ref{resonancetable}, which is in agreement with the width of a resonance being positively correlated with the strength of the resonance \citep{Gallardo2019}. The five strongest resonances in Figure \ref{losstimefig1} correspond to the resonances with highest width in semi-major axis in \citet{Lan2019}, whereas the rest differ in ranking.

All parameters in Equation (\ref{eqnlosstime}) are the same across the disk except those relating to the resonance. The strength of the resonance therefore depends highly on $C_{\rm lib}$, $e_{\rm res}$ and $J_3$. 
Since $T_{\rm loss} \propto (\delta a_{\rm p,lib})^2 \propto e_{\rm res}^{J_3}$, a small change in $e_{\rm res}$ for higher order resonances ($J_3 > 1$) will significantly affect the ranking of the resonances. 

The observed eccentricities $e_{\rm res}$ used to calculate $T_{\rm loss}$ are averages or Gaussian-centered values from observations (see Table \ref{resonancetable}). TNOs are harder to detect the further their semi-major axes and pericenters are from the Sun and so these observational biases make discovering low-eccentricity distant TNOs difficult. In addition, $e_{\rm res}$ for the resonances without constrained values might deviate from the pericenter fit we calculated. As a result, the ranking of resonance strength in Figure \ref{losstimefig1} is an estimate, and varying $e_{\rm res}$ would change the ordering of the resonances. 

We highlight that because we pick the eccentricities of objects observed to occupy each resonance today, we know that those objects have not been lost, and we can use this verified resonance retention to place limits on the planetesimal population driving migration.  However, it may well be that resonances---particularly higher-order resonances---preferentially lost particles at low eccentricities, so that particles are not observed at those eccentricities today.  This behavior generates another pattern that can be explored via comparison with future observational data better able to constrain the low-eccentriciy populations of distant resonances. For example, the third-order 5:2 resonance has a large observed population composed of bodies with large typical eccentricities of 0.4 \citep{Gladman2012,Volk2016}. As illustrated in \citet{Volk2016}, the low-eccentricity population of the 5:2 MMR is not well-constrained given current data.  At observed eccentricity of 0.2, the 5:2 MMR is significantly weaker than at 0.4 (i.e., the migration timescale decreases by $2^{J_3}$ where the order of the resonance, $J_3$, is 3, for a given $\mathcal{M_{\rm LG}}$). Though the resonance width decreases sharply at low eccentricities, a low-eccentricity population in the 5:2 resonance is expected given its similar stable libration zone to the 3:2 and 2:1 MMRs  \citep{Malhotra2018}.  If future observations indicate that the 5:2 MMR does not host a low-eccentricity population, then either (1) it never had one because, for example, the population originated from a planetary gravitational upheaval that scattered planetesimals along trajectories with pericenters in the scattering region and the pericenter distances of low-eccentricity 5:2 resonant objects are too large, or (2) stochasticity during migration caused preferential loss of low-eccentricity objects in the 5:2 MMR.  Conveniently, these two scenarios predict different patterns in the current eccentricities of objects in distant resonaces.  In scenario (1), $e_{\rm res} = 1-q/a_{\rm res}$ for a resonance at semi-major axis $a_{\rm res}$ hosting an object with pericenter distance $q$.  While resonant libration and chaotic diffusion smear observed pericenter distances over time, observed eccentricities in resonance are expected to depend primarily on the semi-major axis of the resonance.  In scenario (2), loss of low-eccentriciy objects due to stochasticity is stronger for higher-order resonances, generating a pattern that depends strongly not only on a resonance's location, but also on its order.  Future characterization of distant resonances at low eccentricities will enable this analysis.  

Changing the eccentricity of the planetesimals near the planet driving migration--- instead of the resonant particle---does not affect the ranking of the resonances.  Instead, the absolute values of the loss times change by a constant factor given a fixed set of disk conditions (this factor is simply a combination of $\mathcal{R}$ and $e_{\rm H}/e$).

The large observed population in the 3rd order 5:2 resonance and the weakness and small population of the 7:3 resonance allow us to place new constraints on planetesimal-driven migration. The timescale and amount of Neptune's migration are confined to a scenario where the weak 7:3 MMR population would not be totally lost and where many objects in the 5:2 MMR would still be retained. Figure \ref{losstimefig1} shows that the 7:4 MMR is actually the weakest, but we do not consider this resonance in our study due to its proximity to the cold classical region and the ease with which it can pick up its neighbors for short timescales \citep{Lykawka2005,Lykawka2007}. As a result, the 7:4 MMR might not be entirely or even dominantly populated by an era of planetesimal-driven migration. We also do not consider the 11:7 and 9:5 MMRs since they do not have constrained population estimates. We discuss constraints due to the 7:3 and 5:2 resonances in Section \ref{sizedist}. This discussion merits renewed attention as we get more data from OSSOS \citep{Bannister2018}, DES \citep{Bernardinelli2022} and the highly anticipated LSST \citep{LSST2009}, at which point the evolution of $e_{\rm res}$ during migration and transiently stuck populations \citep[][]{Yu2018} should also be considered.

\subsection{Retention Fraction}\label{secpkeep}

We now translate our results from Section \ref{seclosstime} into the fraction of initially captured objects that are retained in each resonance as a function of disk properties and the total migration time.  
\citetalias{Murray2006} provide analytical solutions for the retention fraction produced by a disk with uniform planetesimal masses.  The probability that a captured TNO is retained in resonance over the duration $T$ of migration is
\begin{equation}\label{retentioneqn}
P_{\rm keep} = \sum_{n=1}^{\infty} \frac{4}{n\pi}\sin{^3\left(\frac{n\pi}{2}\right)}e^{-\lambda_n T} \,\, ,
\end{equation}
where $\lambda_n = (n\pi)^2D/(2\delta a_{\rm p,lib}^2)$.  The diffusion coefficient describing the cumulative random-walk change in semimajor axis, $\Delta a_{\rm p,T}$ due to many encounters over a total time, $T$ is 
\begin{equation}
D =  \frac{\Delta a_{p,T}^2}{T}\,\, ,
\label{diffeqn}
\end{equation}

\noindent As in all diffusive processes, the diffusion coefficient relates this large-scale behavior to the individual random steps in semi-major axis $\Delta a_{\rm p}$ resulting from single encounters spaced in time by an average $\Delta t \equiv 1/ \dot{\overline{N}}$, where $\dot{\overline{N}}$ is the mean encounter rate given by Equations (\ref{meanNdotnoncross}) and (\ref{meanNdotcross}) for non-crossing and crossing orbits. The diffusion coefficient for the small-scale case is then $D \sim \mathcal{C}_4(\Delta a_{\rm p})^2/\Delta t$. 
For our results to be self-consistent, we expect the coefficient $\mathcal{C}$ in Equation (\ref{RMSeqn}) to be the same for all three regimes since this choice generates smooth transitions at regime boundaries. Equating the large-scale and small-scale diffusion coefficients requires setting $\Delta a_{\rm p,T} = \Delta a_{\rm p}\sqrt{\mathcal{C}_4 (T/\Delta t)}$ equal to  Equation (\ref{RMSeqn}) for each regime and solving for $\mathcal{C}_4$. 
We find that $\mathcal{C}_4 = \Big(\frac{\mathcal{C}}{\mathcal{C}_1}\Big)^2 \frac{2\pi}{81}$, $\Big(\frac{\mathcal{C}}
{\mathcal{C}_{2}}\Big)^2\frac{2\pi}{9}$, and $\Big(\frac{\mathcal{C}}
{\mathcal{C}_{3}}\Big)^2\frac{2\pi}{9}$ for cases (a), (b), and (c), respectively.
Since we found that $\mathcal{C}_1$ is 2.5 times larger than $\mathcal{C}_2$ and $\mathcal{C}_3$ (see Section \ref{secsingleverify}), these results verify that $\mathcal{C}$ is approximately the same for all three regimes. This assures us that the normalization for the micro-scale diffusion coefficient is correct and this process is in fact well modeled by a diffusion process.

\begin{figure}
	\centering
    \includegraphics[width=3.3in]{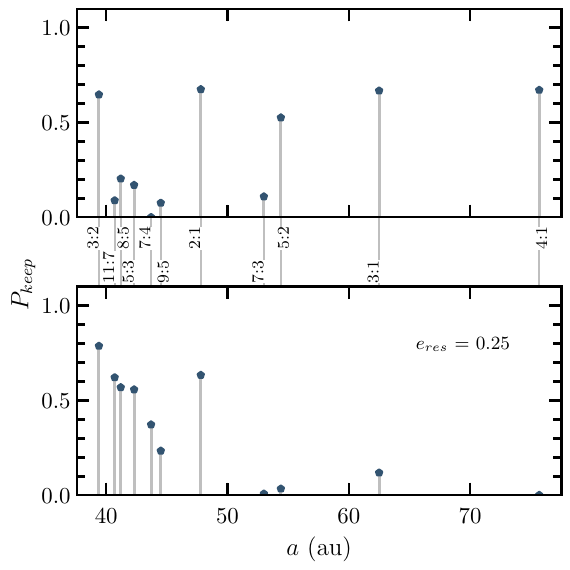}
    \caption{Retention probabilities predicted by Equation \ref{retentioneqn} for Neptune's exterior mean motion resonances (between panels) for the recorded $e_{\rm res}$ (top panel) and $e_{\rm res}=0.25$ (bottom panel).
    For this calculation, $T = 10$ Myr $\mathcal{R} = 1$,  $e = e_H = 0.026$, $a_p = a_d = 26.6$ au, $m = M_{\rm PL}$, $M_{\rm p} = M_{\rm N}$, and $M_* = M_\odot$, and $\mathcal{M} = 0.22$ which corresponds to 2000 Plutos in the planetesimal disk.
    Varying $e_{\rm res}$ changes the strength of the resonances compared to each other, thus highlighting the high dependence on the eccentricity of the objects in MMR in understanding the expected loss from each resonance. 
    Note that if all resonances have a retention fraction of 1, this does not mean that all resonances have equal populations; just that the same proportion is retained.
    }
    \label{pkeepfig}
\end{figure}

In Figure \ref{pkeepfig}, we illustrate the variation in $P_{\rm keep}$ across the resonances we study for the case $e \sim e_{\rm H}$, corresponding to a diffusion coefficient calculated with Equation (\ref{RMSeqn2}). Inputting values that correspond to 2000 Plutos in the primordial planetesimal disk produces varying amounts of loss across the MMRs, with the 7:4 MMR completely obliterated (top panel). Inputting the same values except $e_{\rm res} = 0.25$ (bottom panel) produces different strengths. In particular, the 7:4 MMR becomes stronger than the other weakest resonances for lower $e_{\rm res}$. The 7:3 MMR and in particular the 5:2 MMR produce significantly lower retention fractions for smaller $e_{\rm res}$, highlighting the correlation with $e_{\rm res}$. 
If Neptune traveled in a disk with $e \leq e_{\rm H}$, we expect the same retention fraction behavior since Equations (\ref{RMSeqn1}) and (\ref{RMSeqn2}) are the same when $e = e_{\rm H}$ and $\mathcal{R} = 1$, thus producing the same amount of noise.
For the case where Neptune migrates across a hot disk (i.e. $e>e_H$), the timescale to lose the same fraction of objects as the first two cases increases by a factor of $e^2/e_H^2$.

\begin{figure*}
    \centering
    \includegraphics[width=\textwidth]{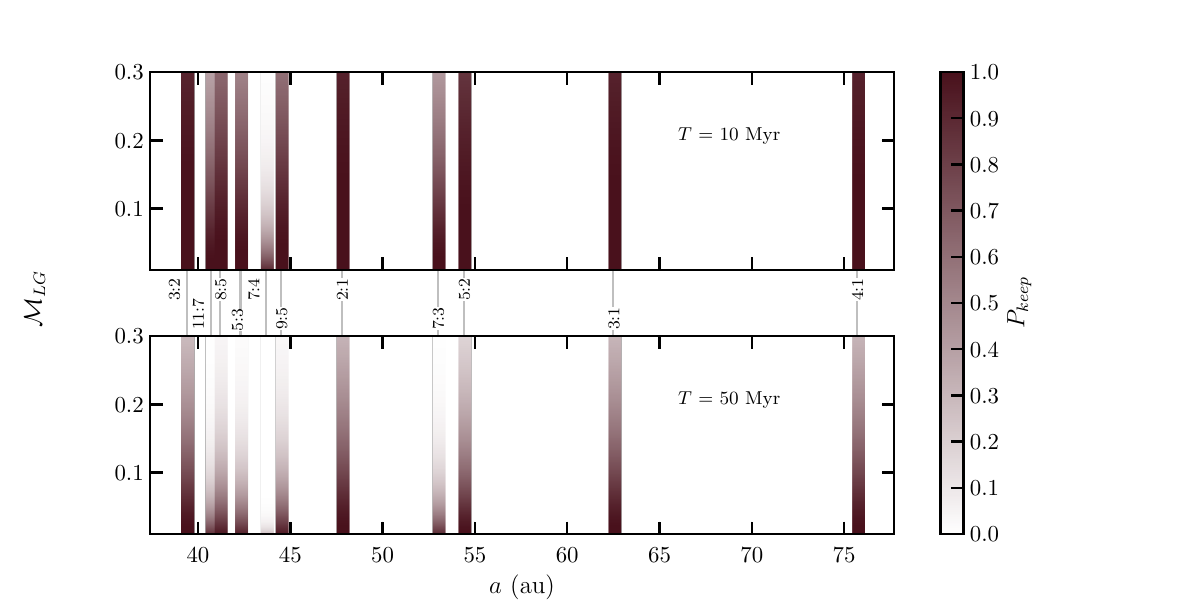}
    \caption{Retention fractions spanning 0 (white) to 1 (dark green) for Neptune's exterior mean motion resonances (between panels) as a function of $\mathcal{M}_{LG}$ for migration timescales 10 million years (top panel) and 50 million years (bottom panel). $P_{\rm keep}$ is calculated with the parameters described in Figure \ref{pkeepfig} (top panel) for a range of surface densities $\mathcal{M}_{LG}$ representing a planetesimal disk with the largest objects having $m = 0.3 m_{\rm PL}$. Increasing the number of large planetesimals in the disk produces more noise, causing a smaller retention fraction (each resonance gradient gets lighter for higher $\mathcal{M}_{\rm LG}$). A similar effect is seen when increasing the migration timescale; more time allows for noise to accumulate, causing more objects to be lost from the resonances.
    }
    \label{fig:pkeepgradient}
\end{figure*}

Increasing the number of large objects in the disk (e.g., increasing $\mathcal{M_{\rm LG}}$ by assuming a more top-heavy size distribution) produces more noise, and fewer objects are retained in resonance (Figure \ref{fig:pkeepgradient}). As the timescale for migration increases from 10 Myr (top panel) to 50 Myr (bottom panel), the total accumulated noise increases, and significantly more loss is seen for a given $
\mathcal{M}_{LG}$.

\section{Influence of Planetesimal Size Distribution}
\label{sizedist}
For a given size distribution of planetesimals, noisiness in the random walk of a planet is dominated by those with maximum $Nm^2$ where $N$ is the number of planetesimals and $m$ is the mass of an individual planetesimal  (c.f. Equation (\ref{RMSeqn}), noting that $N \propto \mathcal{M}/m$). It turns out that larger-radius planetesimals, $s$, produce the highest $Nm^2$. We summarize why that's the case here. The Kuiper Belt size distribution is described by a power law, $dN/ds \propto s^{-q}$, and integrating that gives $N(s) \propto s^{1-q}$. Assuming spherical, uniform-density TNOs, then $N(s)m^2 \propto s^{7-q}$ so that for $q<7$ (typical of Kuiper Belt size distributions) $Nm^2$ is larger for larger $s$. Given that few TNOs have direct size estimates, their measured absolute ($H$) magnitude distribution is typically discussed instead, where $dN/dH \propto 10^{\alpha H}$. The size distribution exponent $q$ is related to $\alpha$ by $q = 5\alpha + 1$.

The number of large planetesimals in Neptune's vicinity at the time of its migration depends both on the total mass in planetesimals at that time and on their size distribution.  The observed mass in Kuiper belt objects \citep[e.g.,][]{Chiang2007} is $\lesssim$1\% of the likely initial mass in solids in the Neptunian region.  Given uncertainties in the evolution of the population of large TNOs, we consider two end-member possibilities: (1) the surface density of TNOs was depleted in a size-independent way, perhaps through dynamical ejection during the epoch of planetary migration, so that the current observed size distribution matches the distribution at the time of migration; and (2) smaller TNOs were cleared from the outer solar system more easily, perhaps through grinding down to dust \citep[e.g.,][]{PanSari2005}, and the large TNOs observed today are the only large objects that ever occupied the migration region.  These possibilities provide and upper limit and lower limit to the expected stochasticity resulting from planetesimal interactions, respectively.

We note that planetesimal-driven migration and damping of the planet's eccentricity result from interactions with all planetesimals and are not dominated by large bodies that house a minority of the mass.  Thus, the stochastic component of the migration cannot be directly inferred from the distance of planetesimal-driven migration or amount of eccentricity damping without knowledge of the planetesimal size distribution.  

\subsection{Current size distribution
}\label{todayssizedist}

Measurements for the absolute magnitude distribution of the Kuiper Belt are often separated into dynamically "cold" and "hot" populations or MMR populations \citep{Fraser2014, Adams2014,Lawler2018,Kavelaars2021,Petit2023}, and recent work demonstrates that a full accounting of each population may require several power law slopes with several breaks \citep{Morbidelli2020}.
We choose the broken power law $H$ distribution for the dynamically hot population provided by \cite{Lawler2018}, to comment on constraints for Neptune's migration, motivated by the idea that the hot population is most likely to have scattered from the planet's vicinity during migration. 

The $H$ distribution for the dynamically hot population from \citet{Lawler2018} has a steep slope ($\alpha = 0.9$) for large objects (low-$H$) and a break magnitude $H_b = 8.3$, corresponding to a radius of 75 km (assuming an albedo, $a=0.04$). Most of the mass resides in planetesimals with sizes near the break.  We calculate the parameterized mass in large objects, $\mathcal{M_{\rm LG}}$, for this size distribution with a simple approximation: 

\begin{equation}
    \begin{aligned}
        &\mathcal{M} M_{\rm disk} = M_{\rm total} = \mathcal{C}s_{b}^{4-q}\\
        &\mathcal{M_{\rm large}} M_{\rm disk} = M_{\rm large} = \mathcal{C}s_{m}^{4-q}
    \end{aligned}
    \label{curlymsizedist}
\end{equation}

\noindent such that

\begin{equation}
    \begin{aligned}
        \mathcal{M_{\rm large}} = \mathcal{M}s_{m}^{4-q}/s_{b}^{4-q}
    \end{aligned}
    \label{curlymfraction}
\end{equation}

\noindent where $\mathcal{M} = 2$ (appropriate for $\sim$ 30 M$_\oplus$ in the disk), $s_b$ is the radius at the break of the broken power law (75 km) and $s_m$ is the maximum radius covered in the distribution (780 km), and $q$ is 5.5 for sizes ranging from the break to the maximum. As a result, $\mathcal{M}_{\rm LG}$ for this size distribution is 0.060, which transaltes to 3\% of the mass in the disk in large (780 km) planetesimals (or 1528 objects)

If the primordial planetesimal disk had the same size distribution as today's hot TNOs, Neptune's noisiness during 80 Myr migration would cause the 7 strongest resonances (as shown in Figure \ref{losstimefig1}) to retain 70-80\% of their resonant objects, while 7:3, 5:3, and 9:5 retain between 30-45\% of objects. The 7:4 MMR loses objects extremely fast compared to other resonances, so we do not include this resonance when discussing "the weakest resonances" further (see Figure \ref{fig:pkeepgradient}). A longer migration timescale produces more loss across the resonances.  For the 7:3 MMR to retain at least $\sim$30\% of its objects, this size distribution results in a maximum migration timescale of 80 Myr.

The retention fraction is highly dependent on the mass of the planetesimal, so we do a similar analysis as described in Equations (\ref{curlymsizedist}) and (\ref{curlymfraction}) for a planetesimal mass $m=m_{\rm pluto}$ and $s_m = 1100$ km. Considering the same size distribution again, now $\mathcal{M_{\rm PL}}=0.036$, which translates to 1.5\% of the total disk's mass in Pluto-sized objects (i.e. 325 objects). For this scenario, the same retention behavior as discussed above occurs, but for a migration timescale of 40 Myrs.

\subsection{Number of Large TNOs}\label{todayslarge}

\begin{figure}
	\centering
        \includegraphics[width=3.3in]{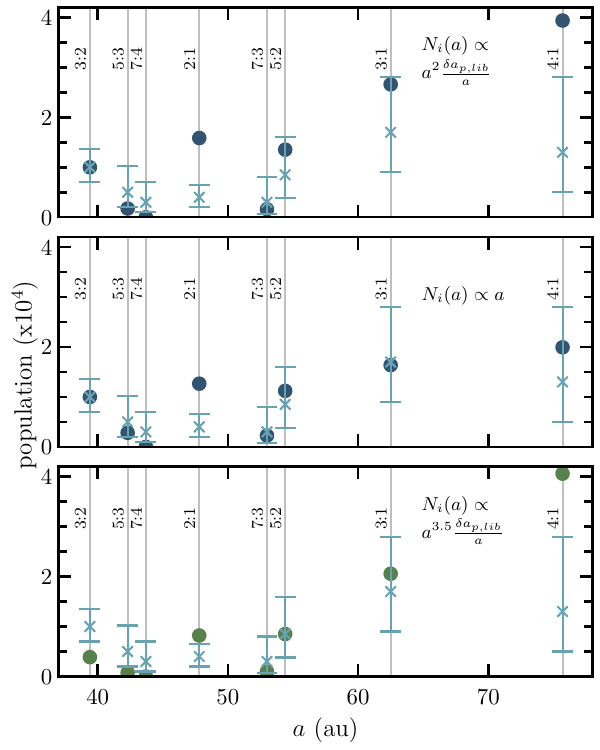}
    \caption{
    The debiased absolute population estimates of resonant TNOs brighter than $H_g=9.16$ and $H_r=8.66$
    (light blue lines with 95\% confidence error bars) provided by \citet{Gladman2012} (5:3, 7:3, 7:4), \citet{Volk2016} (3:2, 2:1, 5:2), and \citet{Crompvoets2022} (3:1, 4:1) compared to noise-reduced populations due to 2000 Plutos around Neptune as it migrates for 10 Myr (circles). 
    We consider two scenarios for filling the mean motion resonances with an initial number of objects $N_i(a)$: (1) a scattering scenario that fills $a-e$ phase space with planetesimals over a region corresponding to orbital pericenters in the scattering region (top and bottom panel) and (2) a substantial amount of ``smooth" planetesimal-driven migration moved Neptune outward, picking up objects into the various resonances (middle panel).
    Scenario (1) requires more work to understand the $a$-dependence of $N_i(a)$, so we consider two distributions ($N_i(a) \propto a^2 \delta a_{\rm lib}/a$, top panel) and ($N_i(a) \propto a^{3.5} \delta a_{\rm lib}/a$, bottom panel), where the first is normalized to the 3:2 estimated population and the latter is normalized to the 5:2 estimated population. See text for further explanation. 
    Scenario (2) results in $N_i(a) \propto a$ which is normalized to the 3:2 estimated population (middle panel).
    }
\label{gladmanfig}
\end{figure}

We switch gears to considering a scenario in which the large TNOs observed today are the only large objects that existed in the disk during Neptune's migration. 
The fraction of large objects, $\mathcal{M}_{LG}/\mathcal{M}$, is simply the fraction of the total mass in large objects over the total mass of the disk. 
Constraints for the timescale and number of large objects in the disk are confined to scenarios where the weakest 7:3 MMR population would not be totally lost given its estimated current population (see Figure \ref{gladmanfig} and Section \ref{sec:obs} for further discussion).
If the disk of planetesimals had the same 8 largest objects from \citet{Brown2008}, corresponding to an average radius $s = 780$ km and $\rho = 2 \rm g/cm^2$, then $\mathcal{M}_{\rm LG} = 0.0003$. This surface density does not cause any loss from the resonances for reasonable migration timescales. To drop a significant fraction of objects in the 7:3 during an epoch of Neptune's planetesimal-driven migration, the disk would need several hundred times more planetesimals with a radius of 780 km, than what exist today.

According to the minimum mass solar nebula model, $10^2-\sim10^{3}$ times more mass existed during early Solar System formation than now \citep{Chiang2007}. If the expulsion of material out of the outer Solar System was due to a mass-independent chaotic, gravitational upheaval, then one can expect up to $\sim10^3$ times more large TNOs than those that exist today. Even in the absence of a dynamical upheaval event, long-term dynamical evolution leads to some depletion of TNOs in MMR over time (see studies for 3:2 and 2:1, e.g., \citealt{Tiscareno2009,Balaji2023}). We comment on the relative retention fractions for long term Nbody effects and Neptune's noisy migration in Section \ref{sec:discussion}.

In addition, there may exist more than 8 large TNOS that simply remain undiscovered. We find the minimum number of large, 780-km objects needed to produce significant noise-induced loss from the MMRs during planetesimal-driven migration. Though only a couple of Pluto-sized TNOs currently are known, we perform this calculation for planetesimals with a Pluto size or twice the size of Pluto (hereafter twice-Pluto) to provide a larger range of constraints. 
A few thousand 780 km objects will produce the similar noise, and thus loss from resonance, as hundreds of Plutos and tens of twice-Plutos. Specifically, the timescale to lose the same fraction of objects for a disk extended from a size $s_{m,1}$ to $s_{m,2}$ will be $(\mathcal{M}_2/\mathcal{M}_1)/(m_1/m_2)$ where the masses $m$ are proportional to $s^3$. 

The migration timescale and number of large objects that would deplete the 7:3 MMR up to 99\%, in strong disagreement with observations, is as follows:
4500 780-km planetesimals with $T = 60$ Myr, 3000 Plutos with $T = 10$ Myr, and 200 twice-Plutos with $T = 10$ Myr. Requiring 25-30 \% retention for the 7:3 MMR produces stronger constraints:
2000 780-km object with $T = 60$ Myr, 1333 Plutos for $T = 10$ Myr, 80 twice-Plutos for $T = 10$ Myr. Note that increasing the number of large objects by 2 and decreasing  the migration timescale by the same factor produces the same results since the argument in the exponential in $P_{\rm keep}$ ($ \lambda_n T$ ) is proportional to $\mathcal{M}T$.

\section{Comparison with Observations}\label{sec:obs}

De-biased absolute population estimates of Neptune's mean motion resonances from \citet{Gladman2012},\citet{Volk2016}, and \citet{Crompvoets2022} provide a good opportunity to directly compare with populations affected by the stochastic model discussed in this paper. These comparisons allow us to comment on the possibility of these resonances' current populations being set during a planetesimal-driven migration era. We consider two methods for filling the mean motion resonances with an initial number of objects $N_i(a)$. The number distribution $N_i(a)$ is determined by the two resonance capture models and respective normalization.  For example, if $N_i(a) \propto a$ and we normalize to the 3:2 estimated population, $N_{\rm{est},3:2}$, such that the predicted population matches the estimated population, then $N_{\rm est,3:2} = N_{i,3:2} P_{\rm{keep}, 3:2}$. Calculating the remaining MMR populations is then a simple relation, which in this case is $N_{i,5:3}=(N_{i,3:2}/a_{3:2})a_{5:3}$. After calculating the initial population for each resonance, we apply the retention fraction calculated in Figure \ref{pkeepfig} to get a population affected by noise during planetesimal-driven migration, $N_{i}(a_{\rm res})P_{\rm keep,res}$  

Scenario (1) appeals to a planetary upheaval that fills the phase space beyond Neptune for a range of pericenters in the scattering region as described in \citet{Balaji2023}, motivated by the simulation results in \citet{Levison2008}.
This pattern results from an early random walk in energy and angular momentum space due to kicks at every perihelion passage with Neptune (or, in principle, with another giant planet, depending on the details of the scattering scenario). The time it takes for Neptune to change an object's angular momentum by an order of itself is proportional to the number of objects at a given semi-major axis for a steady state. \citet{Yu2018} reports that the number of scattering objects at a given semi-major axis from 30 to 100 au is $N_i(a) \propto a^{7/2}$ which they confirm through an empirical fit to simulations from \citet{Kaib2011}. \footnote{Note that \citet{Yu2018} give an incorrect explanation for the a-dependence in their Section 3.2 since the pericenter velocity at high eccentricity and fixed pericenter is not proportional to $a_p$.} On the other hand, \citet{Levison2006} approaches the same problem by looking at planetesimals' random walk in energy due to perihelion passages with Neptune. They find the time it takes to change an object's energy by order of itself is proportional to $a^{-1/2}$, therefore $N_i(a) \propto a^{-1/2}$.
More work is needed to understand the numerical distribution found by \citet{Kaib2011}.  A clue may come from the fact that \citet{Yu2018} find that today's scattering disk of TNOs and today's population of objects transiently-stuck in MMRs comprise a single population, 40\% of which are ``pausing" their random walk in a resonance at any given time. 

In this scenario, we model the relative resonance populations as having populations that follow the background scattering populations' $a$-distribution. The scattering occurs before Neptune reaches its final orbital location so that when this final location is reached, subsets of the scattered objects happen to be in resonant orbits.  
The initial number of objects in a resonance at semi-major axis $a$ is therefore $N_i(a) \propto a^{k}\delta a_{\rm p,lib}/a$, where the number of objects at semi-major axis $a$ in the scattered population scales as $a^k$.
We find that normalizing to the 3:2 MMR population, a power, $k$, from -1/2 to 2, and applying $P_{\rm keep}$, provides a reasonable match to the observed populations, with the 2:1 MMR prediction the most discrepant.  Figure \ref{gladmanfig} (top panel) illustrates the upper end of this range ($k=2$), and $k=3.5$ provides a poor match.   
If we instead normalize to the 5:2 MMR and do not worry about matching the 3:2 MMR population due to complications resulting from secular resonances \citep{Volk2019},  \citet{Yu2018}'s empirical fit, $N_i(a) \propto a^{3.5}\delta a_{\rm p,lib}/a$ provides a more reasonable fit compared to the same $k$ normalized to the 3:2 estimated population (green circles, bottom panel).
The 7:4 fraction is close to 0, but this resonance tends to capture nearby objects easily for short periods of time \citep{Lykawka2005,Lykawka2007}.

Scenario (2) represents efficient, preferential capture of objects into MMRs due to the ``smooth" average component of outward, planetesimal-driven migration in a disk with a radial gradient of planetesimal number density. 
The number of objects, $N_i(a)$ picked up into a resonance that ends this migration with final semi-major axis $a_f$  can be estimated by $N(a) = 2\pi a \Delta a \Sigma(a)$ where the distance swept by the resonance $\Delta a$ is given by the distance $\Delta a_N$ migrated by Neptune (the same for all resonances) multiplied by $a/a_N$, which stays constant in time for each resonance through Kepler's third law and hence may be evaluated at the final semi-major axes of both bodies.  For a fiducial surface density of planetesimals in the disk prior to resoance capture $\Sigma(a) \propto 1/a$, we find $N(a) \propto a$. Figure \ref{gladmanfig} (middle panel) provides results for a smooth migration as discussed in \citet{Nesvorny2016} with "grainiess" due to 2000 Pluto sized planetesimals migrating over 10 million years.
Other than the 7:4 (which is likely to be repopulated by neighbors; \citealt{Lykawka2007}) and 2:1 MMRs, the retention fractions reproduce observed population estimates. 

The results in Figure \ref{gladmanfig} are not conclusive, but the fits are promising.  Data from the Rubin telescope's LSST survey is expected to constrain resonance populations for a much larger number of resonances.  With these new data, it will be possible to use patterns of resonance occupation to determine whether stochasticity did---or equally interestingly whether it didn't---play a significant role in shaping the dynamical structure of the trans-Neptunian region.

\section{Discussion}\label{sec:discussion}
The results presented in Section \ref{sec:obs} are promising as we await unprecedented data from LSST. Better observations will lead to improved constraints for the relative populations and eccentricities of the objects in resonance. Such data will merit revisiting this analysis and improving on our assumptions. We discuss future work and additional considerations below.

\subsection{Comparing the Imprint from Neptune's Noisy Migration and Long Term Evolution}
\label{longvsnoisy}

We have calculated relative retention fractions for ten of Neptune's exterior mean motion resonances for the case of Neptune's \textit{noisy} planetesimal-driven migration lasting 10 or 50 Myr. Between the end of such an epoch of migration and now, additional small-scale perturbations imparted onto TNOs by the planets cause TNOs in MMR to be lost from resonance. 
To determine whether the pattern of resonance population we aim to match in this work (Figure \ref{gladmanfig}) indeed arises from noisy early migration, it will be necessary to understand the pattern of retention across resonances produced by this long-term dynamical sculpting.
\citet{Tiscareno2009} and \citet{Balaji2023} studied the effect of small scale perturbations over billions of years between the giant planets and objects in 3:2 and 2:1 MMRs. 
\citet{Tiscareno2009} quotes retention of 37\% (27\%) and 24\% (15\%) after 1 Gyr (4 Gyr) for the 3:2 and 2:1, respectively. For the 3:2 MMR long term evolution simulaitons from \citet{Balaji2023}, 29\% (19\%) were retained after 1 Gyr (4.5 Gyr). 
\footnote{\citet{Tiscareno2009}'s 3:2 initial population are tighter in resonance than \citet{Balaji2023}'s, so a more fair comparison is possible by finding the retention fraction relative to the 10 Myr population in \citet{Balaji2023}'s simulations. If we count the resonant objects at 10 Myr as the initial population, thus more tight in resonance, then 39\% (25\%) remain in resonance after 1 Gyr (4.5 Gyr), providing a closer match to \citet{Tiscareno2009}. The retention percentages in the text were not quoted in \citet{Balaji2023}---we calculated them based on simulation data from \citet{Balaji2023}. } 
Both studies get comparable magnitudes for the retention fractions, but relative retention fraction across a larger collection of MMRs has not been systematically explored. 
Comparing the relative population estimates across the resonances, rather than the percentages directly, will provide insight into which method left a larger impact: long term evolution or noisy migration. Future population estimates and studies similar to this work and \citet{Tiscareno2009,Balaji2023} extended to a larger range of MMRs will help constrain which method had a larger role in shaping the structure in the outer Solar System.

\subsection{Retention Fractions in the Presence of a Ninth Planet}
We do not include the presence of a hypothetical Planet Nine in our analysis to calculate the retention fractions of objects in MMR between 39 and 75 au. \citet{Clement2021} considered the impact of an undetected planet around $\sim 500$ au on distant $n:1$ resonances between $\sim$60-180 au (3:1-14:1). In their study, the farthest resonances we study (3:1 and 4:1), are similarly eroded in simulations with and without Planet Nine. In addition, TNOs with perihelia less than around 40 au are expected to remain coupled with Neptune \citep{Clement2021}. Therefore, we expect the relative retention rates from this work to not be affected from an undetected distant planet. Future work similar to \citet{Clement2021} and that done here for even more distant $n:1$ resonances that are significantly perturbed by a planet at 500 au might offer an interesting way to constrain the Planet Nine hypothesis. 

\subsection{Equivalent Analysis with Inclination-Type Resonances}
The noisy migration framework in this paper was originally derived for the first order 3:2 MMR with Neptune, where the resonance width (Equation (\ref{alibeqn})) and $C_{\rm lib}$ are derived for low eccentricities to first order. In this paper, we have shown that higher order, weaker resonances, provide strong constraints for Neptune's migration given their expected smaller population and relative retention fraction. To second and third order, additional inclination and mixed eccentricity/inclination weak resonances exist that we have not considered here (See Chapters 6 and 8 in \citet{Murray1999}). An analysis of these resonances analogous to that presented here would be interesting. We leave this for future work, once LSST provides a larger amount of data to constrain eccentricities, inclinations, and population estimates. 

\subsection{Neglecting Planetesimal-Planetesimal Interactions}
\label{nomutualgravity}
We do not include gravitational interactions between planetsimals in the simulations described in this work in an effort to reduce computation time. We justify this choice here.
Including the self-gravity of the disk affects the dynamics of a planetary system in a number of ways: (1) a planet's orbit will precess due to the disk's effect on the planet, (2) the planetesimals' relative eccentricities will be both excited and damped through mutual gravitational interactions, thus changing the dynamics of the system, and (3) the total migration of a planet will be affected. For (1), it is not obvious that the planet precessing would cause a significant change to the random walk of the planet and thus loss of TNOs from resonance since the encounter times between the planet and planetsimals driving the migration are shorter than the precession time. For (2), to reproduce an accurate model for the excitement and damping of planetesimals, we would require collision and gas damping physics which mostly impacts the smallest planetesimals of which there are a larger amount. To accurately describe the relative eccentricities we would need hundreds of thousands of particles with a mass distribution, which is currently computationally infeasible. We leave the relative eccentricities of the planetesimals as a free parameter, shown by the low-eccentricity and high-eccentricity simulation sets in \ref{secmanyverify} and do our analysis for the largest planetsimals since they yield the most noise. For (3), we are only interested in simulating the noisy part of the migration, not the overall behavior. The timescale for the noisy migration component is shorter than the total migration, so it is sufficient to consider these separately. \citet{Fan2017} found that for a Nice model type of dynamical evolution, a self-gravitating disk did not lead to significantly different final planetary orbital behavior compared to a non-self-gravitatating disk. In addition, the planetesimal evolution for both cases did not appear to be significantly distinguishable after an instability was triggered. On the other hand, \citet{quarles2019} found that the self stirring of the disk has a greater impact on the giant planets' dynamics than previously thought. The effect of the self-gravity of a disk on the dynamics of planetary systems is an active area of research where improved computational capabilities such as GPUs are providing an avenue to study this further \citep{Grimm2014,Grimm2022,Kaib2024}

\subsection{Surface Density Profile Effect on Resonance Retention Fraction}

In calculating the relative retention fractions for each mean motion resonance in this work, we assume a constant disk surface density parameterized by $\mathcal{M}$ and by the fiducial semi-major axis of the disk, $a_d$, which we hold constant (see Equation \ref{eqn-sigmam}). We note that the semi-major axis of the planet does not change substantially over the course of our simulations (see Figure \ref{randomwalkfig}), so we are simulating a snapshot in time during the planet's migration.  In reality, the total resonance loss will be produced by noise properties integrated over the full migration.

Consider a disk surface density distribution that follows a negative power law, $\Sigma_m \propto a^{-\beta}$, with $\beta > 0$. From Equation~(\ref{retentioneqn}), $P_{\rm keep}$ declines exponentially when $\lambda_n T \propto a^{1/2-\beta}T$ increases, where $ T $ is the total migration timescale.
Within a given disk, an equivalent amount of time spent at small semi-major axis produces more noise than at large semi-major axis as long as $\beta > 1/2$. 
In other words, for typically-assumed surface density power law distributions with $1 \lesssim \beta \lesssim 3/2$, $a^{1/2-\beta}T$ is largest at small $a$ as long as $T$ remains roughly constant.  As the overall disk surface density decreases with time due to scattering by the planets, the surface density encountered by Neptune as a function of semi-major axis likely declines more rapidly than the initial disk surface density distribution (effectively producing a larger $\beta$).  Thus, the beginning of the epoch of smooth migration (when chaotic planetary motion ceases and $T$ becomes long) is likely to dominate loss of TNOs from resonance due to noise.  The details of this time evolution merit further work.   

\section{Conclusion}\label{sec:summary} 

A planetesimal disk with fixed surface mass density $\Sigma$ generates more stochasticity when composed of larger, fewer planetesimals. An epoch of planetesimal-driven planetary migration, capturing objects into MMR and losing a fraction due to noise, will leave clues behind regarding its dynamic past. The analysis shown in this paper appeals to migration produced by planetesimals over timescales of up to 100 Myr or to scenarios after a gravitational upheaval, where late-stage planetesimal-driven migration is seen in numerical simulations \citep[e.g.,][]{Tsiganis2005} on timescales of up to a few tens of Myr. The weak MMRs lose objects more easily, so we use them to comment on the size distribution of the planetesimal disk such that the weakest resonances with currently observed occupants do not lose their populations due to the stochasticity of planetesimal-driven migration.

In this paper, we tested the validity of modeling the randomness in planetary migration as Brownian motion by numerically confirming results from \citetalias{Murray2006} and deriving coefficients that were left out during order-of-magnitude derivations.
We analytically determine the relative retention fraction of populations of TNOs captured by Neptune into a range of MMRs during migration.  Increasing the mass in large planetesimals contained in the disk or increasing the migration timescale produces more loss across all resonances. However, the proportion of this reduction is not equal across the resonances, with particularly large losses occurring in the 5:3, 7:3, 7:4, and 9:5 resonances. The smallest losses occur in the 2:1, 3:1, 4:1, and 3:2 resonances.  The `strength' of resonances can be ranked in terms of their loss times and retention fractions (see Figures \ref{losstimefig1} and \ref{pkeepfig}). We note that increasing the eccentricity of the planetesimal disk increases the timescale needed to produce the same amount of loss by a factor of $e^2/e_H^2$ for planetesimal eccentricities larger than $e_H$. In addition, changing the eccentricity of the resonance from $e_{\rm res,1}$ to $e_{\rm res,2}$ changes the loss time by a factor of $\big(e_{\rm res,2}/e_{\rm res,1}\big)^{J_3}$, where $J_3$ is the order of the resonance.

We apply our analysis to the case of the Solar System with a planetesimal disk twice the mass of Neptune, corresponding to $\mathcal{M} =2$.
We considered size distribution end-member cases accounting for the mass of large planetesimals with maximum radii $s_m = 780$, 1100, and 2200 km. If part of the 7:3 MMR's population was obtained before or during the epoch of planetesimal-driven migration, it cannot have a retention fraction of zero. Given this condition, we find four conditions that must have been met for a given maximum size of planetesimal to retain at least 30 \% of TNOs in 7:3 MMR for a disk with $e \le e_H$: 

\begin{itemize}
    \item the size distribution with $q = 5.5$ extended to a maximum radius of 780 km requires $\lesssim$80 Myr migration 
    \item the size distribution with $q = 5.5$ extended to a maximum radius of 1100 km requires $\lesssim$40 Myr migration
    \item 1333 Plutos require a 10 Myr migration
    \item 80 twice Pluto-size objects require $\lesssim$10 Myr migration
\end{itemize}

Additional scenarios may be extrapolated using these results using the facts that for constant $\mathcal{M}T$ and constant $(\mathcal{M}_2/\mathcal{M}_1)/(m_1/m_2)$, the same level of resonance retention is produced. A more massive disk ($\mathcal{M} > 2 $) would produce constraints for shorter timescales.  Observational constraints on the population and size distribution of bodies with radii $\gtrsim$780km are limited by small number statistics, leaving uncertainty in the expected early population and sizes of large bodies.  In the limiting case that the large bodies currently observed in the TNO region are the only large planetesimals that ever existed in the outer disk, planetesimal-driven migration is effectively smooth, with stochasticity producing no meaningful resonace loss.
The unprecedented data for TNOs from the Vera C. Rubin Observatory will improve these parameters, and a similar analysis will cover the populations of a significantly larger collection of distant resonances, producing stronger constraints.

The values stated above that decrease the 7:3 population by 70\% would decrease the 5:2 by 25 \% only.
Observations show a large population in the  5:2 resonance---comparable to the number in the 3:2 MMR, with high eccentricities ranging from approximately 0.2 to 0.6 \citep{Gladman2012,Volk2016,Malhotra2018}.  Presumably, if the 5:2 resonance was populated before or during the epoch of planetesimal-driven migration, a large proportion of the captured population must have been retained. This large proportion of high eccentricity bodies could be due to detection biases since beyond 46AU, resonant bodies with low eccentricities are difficult to detect. As noted in Section \ref{todayssizedist}, the 5:2 resonance is relatively weak for $e_{\rm res} = 0.25$ but substantially stronger for $e_{res} = 0.4$, the typical value for observed 5:2 objects measured (Figure \ref{pkeepfig}). 
If future surveys sensitive to fainter objects find many low-eccentricity objects in the 5:2, the low eccentricity 5:2 will generate constraints similar to or better than those from the 7:3. The Minor Planet Center\footnote{\url{https://www.minorplanetcenter.net/data}} shows that a few objects at low eccentricities have been found, but not through a survey that allows a debiased population estimate.

Future investigations of migration scenarios, resonance capture efficiency, and unprecedented data from the Vera C. Rubin Observatory will allow for direct comparison between resonance retention fractions and improved population estimates. We expect decent population estimates for mean motion resonances up to 200 au; with this data, we will be able to investigate the trends found in Figure \ref{gladmanfig}. In particular, patterns in resonance population that depend on the order of resonances as calculated in this work will be a strong indicator that stochasticity played a significant role in sculpting the trans-Neptunean, region.  Conversely, the lack of such patterns would indicate that this random process did not shape the populations we observe today.

\section*{Acknowledgements}

We thank Kat Volk for their comments on an earlier draft of this paper and the anonymous reviewer for their comments which led to an improvement in this paper and addition of Section \ref{sec:discussion}.
AHR thanks the LSSTC Data Science Fellowship Program, which is funded by LSSTC, NSF Cybertraining Grant \#1829740, the Brinson Foundation, and the Moore Foundation; her participation in the program has benefited this work.
This material is based upon work supported by the National Science Foundation Graduate Research Fellowship Program under Grant No. 1842400. Any opinions, findings, and conclusions or recommendations expressed in this material are those of the author(s) and do not necessarily reflect the views of the National Science Foundation. AHR and RMC acknowledge support from NASA grant number 80NSSC21K0376.
We acknowledge use of the lux supercomputer at UC Santa Cruz, funded by NSF MRI grant AST 1828315. 

\section*{Data Availability}

The code for MERCURY that were used to produce the simulation data and plots can be found at \url{https://github.com/ahermosillo/stochastic-randommigration}.

\bibliographystyle{mnras} 
\bibliography{ms}

\appendix
\section{Average Migration}\label{sec:appendix}
The average component of planetesimal-driven migration in the outer solar system is determined by global asymmetries set up as planetesimals are transeferred between the gravitational zones of influece of the various giant planets \citep{Fernandez1984}.  Since the model presented here only includes one planet, this migration is not reproduced.  We show here that the average migration seen in our simulations may be derived using the same analytic framework that we use to model stochasticity, verifying the self-consistency of our results.

The mean component of planetesimal driven migration may be thought of as resulting from a form of dynamical friction.  Regardless of planetesimal eccentricity, a population of planetesimals with semi-major axis deviation $x = a - a_p$ shears past the planet with bulk relative velocity $v_{\rm rel} \approx -(3/2) \Omega_p x$.  Thus, the dynamical friction felt by the planet from planetesimals with $x = x_0$ and those with $x = -x_0$ roughly cancels.  Migration results from the small difference felt between the torque from planetesimals at positive and negative $x$. The difference in relative velocity on either side of the planet ($x = \pm x_0$) due to disk curvature is of order $a_p \left(\frac{d^{2}\Omega}{da^2}\right)2x_0 \sim (15/2)\Omega_p x (x/a_p)$.  Power-law variations in other quantities on either side of the planet similarly lead to differences of order the single-sided torque multiplied by $x/a_p$.  This second-order behavior is why the average distance migrated in Figure \ref{histfig} is not much larger than the standard deviation.  

Considering planetesimals with a single value of $x$, the average (one-sided, Sun-centered) torque experienced by the planet is $\tau_1 \approx a_p M_p \Delta v_{\parallel, p} \dot{\overline{N}}$, where $\dot{\overline{N}} \approx (\Sigma_m/m) \Omega_p x^2$ is the average rate at which the planetesimals shear past the planet at distance $x$.  The change in orbital velocity of the planet per encounter in the aximuthal direction is $\Delta v_{\parallel,p} \sim -(m/M_p) \Delta v_{\parallel}$ where $\Delta v_{\parallel}$ and $\Delta v_{\perp}$ are the velocity changes experienced by the planetesimal in the azimuthal and radial directions, respectively, resulting from a single encounter.  From the impulse approximation, $\Delta v_{\perp} \sim GM_p/(xv_{\rm rel})$ and in the limit that tidal gravity from the star does not change the energy of the planetesimal substantially over the encounter,
\begin{equation}
\Delta v_{\parallel} \approx 
\begin{cases}
\Delta v_{\perp}^2/v_{\rm rel} & \text{, $e < \Delta v_{\perp}/(\Omega_p a_p)$,} \\
 v_{\rm ep} \Delta v_{\perp}/v_{\rm rel} & \text{, $e > \Delta v_{\perp}/(\Omega_p a_p)$}
\end{cases}
\end{equation}
where $v_{\rm ep} \sim e\Omega_p a_p$ is the particle's epicyclic velocity so that
\begin{equation}
\Delta v_{\parallel} \sim 
\begin{cases}
G^2 M_p^2/(x^2 v_{\rm rel}^3) & \text{, $e < e_H^3(a/x)^2$,} \\
G M_p v_{\rm ep}/(x v_{\rm rel}^2) & \text{, $e > e_H^3(a/x)^2$.}
\end{cases}
\end{equation}
Note that the criteria separating the two cases matches the criteria separating the two cases in Equation (\ref{NCeqn}).

Combining these expressions yields 
\begin{equation}\label{eqn-torque1side}
\tau_1 \approx -\Sigma_m \Omega_p^2 a_p^4 \times \begin{cases}
 (M_p/M_*)^2 (a_p/x)^3  & \text{, $e < e_H^3(a/x)^2$,} \\
e(M_p/M_*)(a_p/x) & \text{, $e > e_H^3(a/x)^2$.}
\end{cases}
\end{equation}  
The small-eccentricity case of Equation (\ref{eqn-torque1side}) is equivalent to 2D dynamical friction resulting from a population that does not extend arbitrarily close to the planet.  This one-sided torque yields a rate of change of the planet's semi-major axis given by the approximation $\dot{a}_p \sim (\dot{L}/L) a_p$, resulting in
\begin{equation}
\frac{da_p}{dt} \sim \frac{\tau_1}{M_p a_p \Omega_p} 
\sim \dot{\overline{N}} \Delta a_{\rm p,n}  
\end{equation}
where $\Delta a_{\rm p,n}$ is given by Equation (\ref{NCeqn}).  

The net torque resulting from the difference between torques from particles at $\pm x$ is $\tau \sim f\tau_1 (x/a_p)$, where $f$ is an order-unity coefficient resulting from the combined powerlaw exponents of all quantities changing in the disk on either side of the planet.  For the first case in equation \ref{eqn-torque1side} that gives
\begin{eqnarray}\label{eqn-torque}
\tau \sim f \Sigma_m \Omega_p^2 a_p^4 \times \begin{cases}
(M_p/M_*)^2 (a_p/x)^2 & \text{, $e < e_H^3(a/x)^2$,} \\
e (M_p/M_*) & \text{, $e > e_H^3(a/x)^2$.}
\end{cases}
\end{eqnarray}
The torque from the lower-eccentricity case of  Equation (\ref{eqn-torque}) is familiar from Type I migration, where $x$ is replaced by $H$ in a gas disk since gas perturbations are limited to move at the sound speed, meaning that the closest coherent Lindbland resonances pile up at $x\approx H$.

We can apply our normalization coefficents to this calculation in the following way. The accumulated encounters influence the average distance migrated, such that $\Delta a_{\rm avg} = f(x/a_p)\overline{N} \Delta a_p$ where $\Delta a_p$ represents the change in semi-major axis due to a single encounter and $\overline{N} = \dot{\overline{N}}T$ is the average number of encounters.  For all eccentricities, this average migration comes from the average bulk relative velocity, due to shear, so that $\dot{\overline{N}} = (\Sigma_m/m) \Omega_p x^2$ (folding any coefficient into the overall coefficient we are calculating).  The random walk distance $\Delta a_{\rm p,T} = (\overline{N})^{1/2} \Delta a_p$ 
so that $\Delta a_{\rm avg} = f(x/a_p)\Delta a_{\rm p,T}^2/\Delta a_p = \mathcal{C}_5 f(x/a_p) \dot{\overline{N}} T \Delta a_{\rm p,n}$
for non-crossing cases 1 and 2, respectively.
Plugging in Equations (\ref{RMSeqn1}) and (\ref{RMSeqn2}) reveals the value of $\mathcal{C}_5$.
For non-crossing case 1, the coefficient is $\mathcal{C}_5 = 2 \pi\mathcal{C}^2 /(81 \mathcal{C}_1^2)$ and $\mathcal{C}_5 = 2 \pi\mathcal{C}^2/(9 \mathcal{C}_2^2)$ for non crossing case 2 (this coefficient applies for all high-eccentricity orbits, including crossing orbits). Plugging in, we get
\begin{eqnarray}
\Delta a_{\rm avg} &=& f\frac{x}{a_p}\frac{2\pi}{9}\mathcal{C}^2\frac{\Sigma_m}{m}\Omega_p x^2 T \nonumber \\ &\times& \begin{cases}
\frac{1}{9\mathcal{C}_1}\frac{mM_{\rm p}}{M_*^2}\frac{a_{\rm p}^6}{x^5} & \text{, $e < e_H^3(a/x)^2$,} \\
\frac{1}{\mathcal{C}_2}\frac{m}{M_*}\frac{a_p^4}{x^3}e & \text{, $e > e_H^3(a/x)^2$.}
\end{cases}
\end{eqnarray}

Calculating $\Delta a_{\rm avg}$ at 1 million years for the simulations plotted in Figure \ref{histfig}, using parameters provided for groups 3 and 5 in Table \ref{groupstable} gives $\Delta a_{\rm avg} =  0.0039 f$ au for the low-eccentricity simulaiton and $\Delta a_{\rm avg} =  0.0026 f$ au for the high-eccentricity simulation. These values match the simulation results $\Delta a_{\rm avg} =  0.041$ au and 0.015 au, respectively, for $f = $10 and 6, respectively.  These values are reasonable (comparable to the 7.5 estimated at the beginning of this appendix) and similar to one-another.  We consider this reasonable verification that our interpretation of our numerical results is self-consistent.

\bsp	
\label{lastpage}
\end{document}